\documentclass[runningheads]{llncs}
\usepackage[T1]{fontenc}

\usepackage{latexsym}
\usepackage{amssymb}
\usepackage{booktabs}
\usepackage{enumitem}
\usepackage{graphicx}
\usepackage{color}
\usepackage{algorithm}
\usepackage{amsmath}
\usepackage{amssymb}
\usepackage{float}  

\usepackage{algpseudocode}
\usepackage{subfigure} 

\usepackage{newfloat}
\usepackage{listings}
\begin{document}
\title{AOAD-MAT: Transformer-based Multi-Agent Deep Reinforcement Learning Model considering Agents' Order of Action Decisions \thanks{This manuscript is an extended version of the work accepted as a short paper at the 26th International Conference on Principles and Practice of Multi-Agent Systems (PRIMA 2025). The Version of Record of this contribution is published in Springer's Lecture Notes in Artificial Intelligence series (LNCS/LNAI), and is available online at https://doi.org/}}
\titlerunning{Agents' Order of Action Decisions-MAT}
\author{
    Shota Takayama \and 
    Katsuhide Fujita
}
\institute{Graduate School of Engineering, Tokyo University of Agriculture and Technology}

\maketitle
\begin{abstract}
Multi-agent reinforcement learning focuses on training the behaviors of multiple learning agents that coexist in a shared environment. Recently, MARL models, such as the Multi-Agent Transformer (MAT) and ACtion dEpendent deep Q-learning (ACE), have significantly improved performance by leveraging sequential decision-making processes. Although these models can enhance performance, they do not explicitly consider the importance of the order in which agents make decisions. In this paper, we propose an Agent Order of Action Decisions-MAT (AOAD-MAT), a novel MAT model that considers the order in which agents make decisions. The proposed model explicitly incorporates the sequence of action decisions into the learning process, allowing the model to learn and predict the optimal order of agent actions. The AOAD-MAT model leverages a Transformer-based actor-critic architecture that dynamically adjusts the sequence of agent actions. 
To achieve this, we introduce a novel MARL architecture that cooperates with a subtask focused on predicting the next agent to act, integrated into a Proximal Policy Optimization based loss function to synergistically maximize the advantage of the sequential decision-making.
The proposed method was validated through extensive experiments on the StarCraft Multi-Agent Challenge and Multi-Agent MuJoCo benchmarks. The experimental results show that the proposed AOAD-MAT model outperforms existing MAT and other baseline models, demonstrating the effectiveness of adjusting the AOAD order in MARL.

\end{abstract}

\section{Introduction}
Multi-agent reinforcement learning (MARL) has emerged as a promising approach for solving complex cooperative tasks in various real-world scenarios, including traffic control, robotics, and multiplayer games \cite{marl-book2024,pmlr-v97-jaques19a,lazaridou2017multiagent,Lowe17}. However, MARL faces unique challenges that are beyond those encountered in single-agent reinforcement learning \cite{Rein}, particularly in terms of the non-stationarity of the environment and the exponential growth of the joint action space as the number of agents increases.
Recent advancements in MARL have led to the development of frameworks such as centralized training with decentralized execution (CTDE) \cite{CTDE} and centralized training with centralized execution (CTCE). These frameworks have enabled significant progress in addressing the complexities of multi-agent systems. Within these frameworks, various algorithms have been proposed, including value decomposition methods like QMIX \cite{QMIX} and policy gradient methods like MAPPO \cite{MAPPO}.
A significant breakthrough in MARL was achieved with the introduction of the Multi-Agent Transformer (MAT) \cite{MAT}, which leveraged the power of sequence modeling techniques to enhance MARL performance. By treating the multi-agent decision-making process as a sequence generation task, the MAT model could effectively capture inter-agent dependencies and improve overall team performance. The success of MAT has provided new possibilities for applying advanced natural language processing (NLP) techniques to MARL problems.
Another significant contribution to the field was made by the ACtion dEpendent deep Q-learning (ACE) \cite{ACE} algorithm, which focused on addressing the non-stationarity problem in MARL. ACE introduced a bidirectional action-dependency approach, formulating the multi-agent decision-making process as a sequential one. This formulation allowed for more accurate estimation of individual action values and facilitated more effective cooperative learning.
Although these approaches have shown promising results, they have not explicitly considered the effectiveness of the order in which agents make decisions. The order of action decisions can significantly influence the overall performance and stability of MARL systems. This is particularly important in scenarios where agents have different capabilities or when environment dynamics favor certain action sequences.

To address this issue, we propose an Agent Order of Action Decisions-MAT (AOAD-MAT), a novel Transformer-based multi-agent deep reinforcement learning model that explicitly incorporates and learns the optimal order of agent action decisions. The architecture of the proposed AOAD-MAT model is inspired by the successful MAT architecture and introduces a dedicated mechanism to predict and optimize the sequence in which agents should act.
The proposed architecture considers a subtask focused on predicting the next agent to act on in the learning process to achieve the main task. The subtask is seamlessly incorporated into a Proximal Policy Optimization (PPO) \cite{PPO} based loss function, resulting in a synergistic effect that maximizes the advantage of sequential decision-making. By doing so, the proposed AOAD-MAT learns effective policies for individual agents and optimizes the order in which these policies are executed.

 It is worth noting that the action decision order can improve performance and learning stability in MARL scenarios. To validate this, we conducted extensive experiments on challenging benchmarks, including the StarCraft Multi-Agent Challenge (SMAC) \cite{SMAC} and Multi-Agent MuJoCo (MA-MuJoCo) \cite{MA-MuJoCo} environments. The experimental results show that the proposed AOAD-MAT consistently outperforms existing state-of-the-art models, including the original MAT, for various scenarios.
Furthermore, we validated the effectiveness of the action decision order on the learning dynamics and stability of MARL systems. Our findings provide insights into how the order of agent actions influences the overall team performance and learning process efficiency.

The main contributions of this paper are summarized as follows.
\begin{itemize}
  \item We propose AOAD-MAT, a novel Transformer-based MARL model that explicitly learns and optimizes the order of agent action decisions.
  \item We propose an order-aware policy learning approach within the PPO framework that incorporates a subtask for next-agent prediction to enhance advantage function maximization.
  \item We evaluated the effectiveness of the proposed AOAD-MAT through comprehensive experiments on the SMAC and MA-MuJoCo benchmarks, demonstrating its superior performance compared to existing methods.
  \item We provide insights into the importance of the order of action decisions in MARL and propose new research directions.
\end{itemize}

\section{Related Works}
Early MARL approaches were primarily categorized into fully independent and centralized methods. Independent Q-learning \cite{IQL} pioneered the extension of single-agent reinforcement learning to multi-agent settings by applying Deep Q-Network \cite{DQN} to each agent independently. However, this approach often struggled with the non-stationarity inherent in multi-agent environments.
Although theoretically promising, fully centralized methods have received less attention due to scalability issues arising from the exponential growth of joint action spaces. Some centralized methods have attempted to mitigate this issue through information exchange mechanisms. However, these centralized methods often fall short of the expected performance improvements compared to other methods.
The centralized training with decentralized execution (CTDE) paradigm has emerged as a compromise between fully independent and centralized methods. There are two main categories of algorithms in the CTDE framework: value factorization and policy gradient \cite{PG} methods.

Value factorization methods, such as Value-Decomposition Networks (VDN) \cite{VDN} and QMIX \cite{QMIX}, approximate the joint value function as a combination of individual value functions. These methods often adhere to the Individual-Global-Max (IGM) \cite{QTRAN} principle. However, these methods face challenges of mismatches between optimal joint and individual value functions during training, leading to decreased sample efficiency.
The PG methods in CTDE have gained significant advancements, particularly with the adaptation of trust-region methods like Trust Region Policy Optimization (TRPO) \cite{TRPO} and PPO \cite{PPO}. Independent PPO (IPPO) \cite{IPPO} and Multi-Agent PPO (MAPPO) applied PPO to multi-agent settings, whereas Heterogeneous-Agent PPO 
(HAPPO) \cite{HAPPO} highlighted the importance of sequential agent updates, and A2PO \cite{a2po} further investigated the impact of the update order of agents during training. Additionally, some approaches \cite{DAG,GCS} have leveraged graph structures among agents to infer effective action sequences based on inter-agent dependencies.

A significant breakthrough in MARL was achieved with the introduction of sequence modeling techniques in the centralized training using the centralized training with centralized execution (CTCE) framework. These methods effectively reframe multi-agent scenarios as single-agent problems to address the scalability issues of earlier centralized approaches. MAT \cite{MAT}, an example of the CTCE approach, employs a Transformer-based architecture to model inter-agent interactions and facilitate sequential agent learning. This approach demonstrated improved performance by capturing the complex dependencies between agents' actions and treating the multi-agent problem as a sequence prediction task.

Similarly, ACE (Cooperative Multi-agent Q-learning with Bidirectional Action-Dependency), another example of the CTCE method, addresses the non-stationarity problem by introducing bidirectional action dependency \cite{ACE}. By modeling the multi-agent scenario as a sequential decision-making process, ACE effectively reduced the multi-agent problem to a single-agent setting, thereby enhancing cooperative learning in multi-agent environments.
A recent and notable contribution to MARL is the formation-aware exploration (FoX) model \cite{FOX} the addresses the challenge of efficient exploration in partially observable multi-agent environments. FoX introduces the concept of formation to define state equivalence relationships, which effectively reduces the exploration space and improves sample efficiency. 
Although FoX focuses on enhancing exploration efficiency through spatial relationships, our proposed AOAD-MAT takes a different approach by focusing on the order of agent action decisions. The proposed AOAD-MAT, builds upon these foundational concepts by explicitly modeling the order of agent action decisions in the CTCE framework. Unlike FoX, which reduces exploration space, our proposed AOAD-MAT improves exploration capability by considering the sequence of agent decisions. Specifically, the proposed AOAD-MAT model introduces a dual prediction mechanism: it predicts the next agent to act (a discrete output) and the action of that agent (which can be either discrete or continuous, depending on the environment).
This leads us to the consideration of hybrid action spaces in MARL. Hybrid PPO (H-PPO) \cite{H-PPO} is an effective approach for handling continuous and discrete action outputs in a single architecture. Although H-PPO is not a MARL method, it is particularly important in environments with complex action spaces, such as those found in some MARL scenarios, and shares some conceptual similarities with our AOAD-MAT's ability to handle diverse action types.

The proposed AOAD-MAT, builds upon these foundational concepts by explicitly modeling the order of agent action decisions in the CTCE framework. While not directly dealing with hybrid action spaces like H-PPO or formations like FoX, the proposed AOAD-MAT model handles discrete and continuous probability distributions in its output and focuses on the sequential aspect of agent interactions. 
By integrating this novel aspect into the Transformer-based architecture, we demonstrate the significance of action order in multi-agent cooperation and its effectiveness on overall performance and learning stability, while maintaining the advantages of centralized execution. 

\section{Preliminaries}
\subsection{Multi-Agent Reinforcement Learning (MARL)}
The cooperative MARL scenario is modeled as an extension of the Markov decision process, known as a Markov game \cite{MAPG}. This framework is defined by the tuple $\langle \mathcal{N}, \mathbb{O}, \mathbb{A}, R, P, \gamma \rangle$. $\mathcal{N} = \{1, \ldots, n\}$ denotes the set of agents, $\mathbb{O} = \prod_{i=1}^{n} \mathcal{O}_i$ represents the joint observation space with $\mathcal{O}_i$ being the local observation space of agent $i$, and $\mathbb{A} = \prod_{i=1}^{n} \mathcal{A}_i$ denotes the joint action space where $\mathcal{A}_i$ is the action space of agent $i$. 

The bounded joint reward function is defined as $R: \mathbb{O} \times \mathbb{A} \rightarrow [-R_{\text{max}}, R_{\text{max}}]$, while $P: \mathbb{O} \times \mathbb{A} \times \mathbb{O} \rightarrow \mathbb{R}$ defines the transition probability function. Finally, $\gamma \in [0, 1)$ denotes the discount factor.

At each time step $t \in \mathbb{N}$, every agent $i \in \mathcal{N}$ perceives an observation $o_i^t \in \mathcal{O}_i$ and executes an action $a_i^t$ based on its policy $\pi_i$, which is the $i$th component of the agents' joint policy $\pi$. 
The joint observation is denoted as $\textbf{o} = (o_1, \ldots, o_n)$. All agents act concurrently, and their collective actions determine the team's behavior. 

The transition function $P$ and joint policy induce an observation distribution $\rho_\pi(\cdot)$.
Following each action, the team receives a joint reward $R(\textbf{o}_t, \textbf{a}_t)$, and the environment transitions to a new state, which is observed as $\textbf{o}_{t+1}$ according to $P(\cdot | \textbf{o}_t, \textbf{a}_t)$. This process continues indefinitely, with agents accumulating a discounted cumulative reward of $R^\gamma = \sum_{t=0}^{\infty} \gamma^t R(\textbf{o}_t, \textbf{a}_t)$.

\subsection{Multi-Agent Advantage Decomposition}
Recent MARL approaches such as MAPPO \cite{MAPPO} and HAPPO \cite{HAPPO} introduced the Multi-Agent Advantage Decomposition Theorem, which provides a foundation for decomposing the joint value function in multi-agent systems. This decomposition addresses the credit allocation challenge \cite{Cre}, where individual agents struggle to discern their specific contributions to the team's performance.
The advantage function is defined as $A_{\pi}^{i_{1:m}}(o, a^{j_{1:h}}, a^{i_{1:m}})$, which quantifies the relative benefit of the joint action $a^{i_{1:m}}$ taken by agents $i_{1:m}$, given that agents $j_{1:h}$ have already acted with $a^{j_{1:h}}$. This formulation allows us to analyze inter-agent interactions and decompose the joint value signal, thereby mitigating the credit allocation problem. The concept is formalized as follows:

\noindent\textbf{Theorem 1 (Multi-Agent Advantage Decomposition)}
\label{Multi-Agent Advantage Decomposition}
\textit{~For any permutation of agents $i_{1:n}$, joint observation $o \in \mathcal{O}$, and joint action $a = a^{i_{1:n}} \in \mathcal{A}$, the following equality holds without additional assumptions:}
\begin{equation}
    A_{\pi}^{i_{1:n}}(o, a^{i_{1:n}}) = \sum_{m=1}^{n} A_{\pi}^{i_{1:m}}(o, a^{i_{1:{m-1}}}, a^{i_m}).
\end{equation}
Theorem (1) suggests an approach for incrementally improving collective behavior. Specifically, if agent $i_1$ selects an action $a^{i_1}$ with a positive advantage $A^{i_1}_{\pi}(o, a^{i_1}) > 0$, and subsequently, for all $k=2, \ldots, n$, agent $i_k$ (aware of its predecessors' joint action $a^{i_{1:k-1}}$) chooses an action $a^{i_k}$ with a positive advantage $A^{i_k}_{\pi}(o, a^{i_{1:k-1}}, a^{i_k}) > 0$, then Theorem 1 guarantees that the team's joint action $a^{i_{1:n}}$ will have a positive overall advantage.

\subsection{Multi-Agent Transformer Approach}
Transformer model \cite{Transformer}, originally developed for NLP tasks, such as translation and text generation, has demonstrated remarkable versatility across various domains. Its effectiveness in sequence modeling has led to its adoption in fields beyond NLP, including image and signal processing. The key strength of the Transformer lies in its attention mechanism, which learns to focus on relevant parts of the input data, thereby enabling powerful representational capabilities.

\begin{figure}[tb]
    \centering
    \includegraphics[keepaspectratio, width=0.8\linewidth]{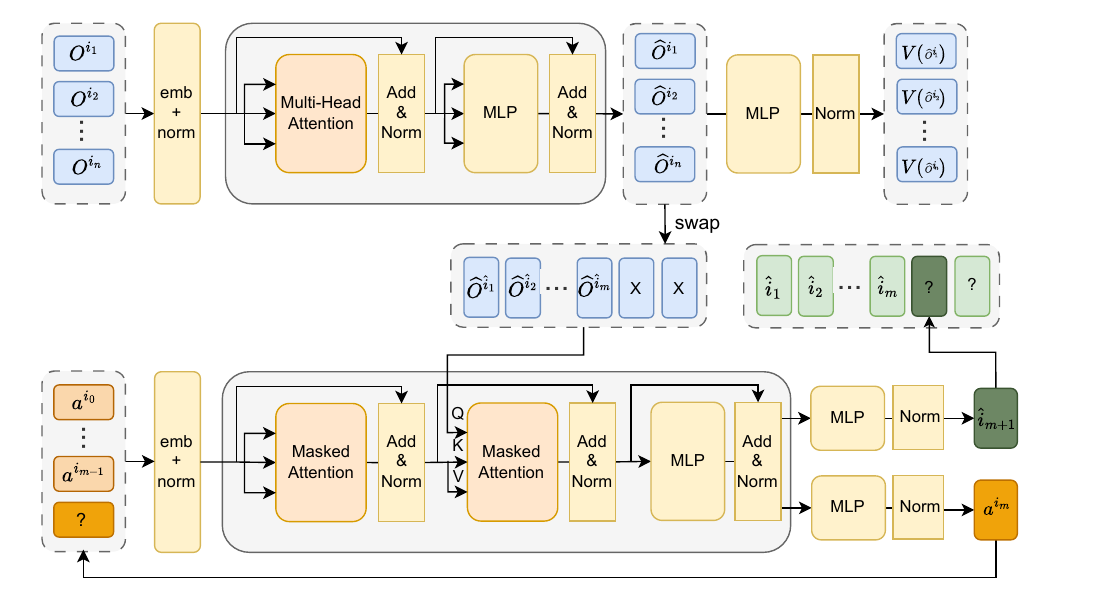}
    \caption{AOAD-MAT's encoder--decoder architecture. The encoder represents the critic network, whereas estimates the state value, while the decoder represents the actor network, which predicts the action and next agent to act.}
    \label{aoad-mat-architecture}
\end{figure}

The application of the Transformer architecture to multi-agent systems stems from the observation that generating an agent's observation sequence $(o^{i_1}, ..., o^{i_n})$ and action sequence $(a^{i_1}, ..., a^{i_n})$ can be framed as a sequence modeling task, which is analogous to machine translation. According to Theorem 1, the action $a^{i_{m}}$ is dependent on the decisions $a^{i_{1:m-1}}$ of all preceding agents. Therefore, the proposed MAT design incorporates an encoder that learns a representation of joint observations and a decoder that generates an output for each agent in an autoregressive manner.

\begin{figure*}
    \centering
    \includegraphics[keepaspectratio, width=0.9\linewidth]{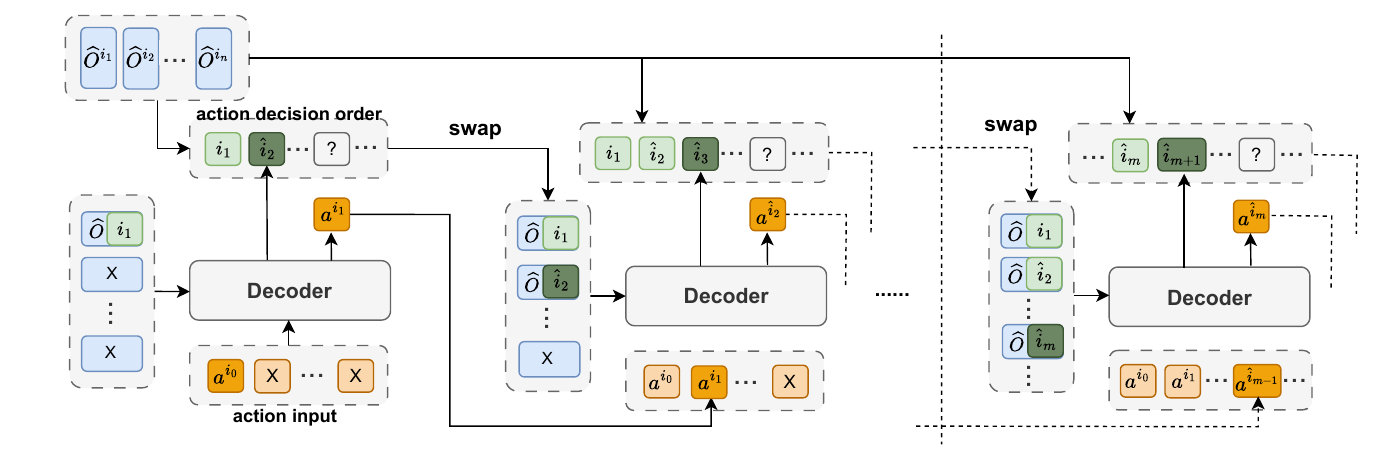}
    \caption{Sequential Action Decision Order Prediction: First, a dummy action $a^{i_{0}}$ is input as the 0th action. At this point, it is necessary to predetermine $i_1$ the agent number that will act first. Then, output $\hat{i}_2$ the agent that will act next. Based on the updated agent action decision order, the latent representations of the observations $\hat{o}$, which are the output of the encoder, are swapped. Finally, the previous action output is added to the input and the same steps are repeated.}
    \label{sequential-action-decision-order}
\end{figure*}

\section{AOAD-MAT: Agent Order of Action Decisions-MAT}
\subsection{Sequential Action Decision Order Prediction}
The input of the encoder, the agent's observation sequence, corresponds to the latent representation of the observations $(\hat{o}^{i_1}, \ldots, \hat{o}^{i_n})$ for each agent. Based on $(\hat{o}^{i_1}, \ldots, \hat{o}^{i_n})$, the decoder outputs the action sequence $(a^{i_1}, \ldots, a^{i_n})$ in a serial manner. The latent representations of the observations can be reordered in an arbitrary order as follows:
\begin{align} \nonumber
&\hat{o}_{\text{swap}} = \gamma(\hat{o}^{i_1}, \ldots, \hat{o}^{i_n}) = (\hat{o}^{\hat{i}_1}, \ldots, \hat{o}^{\hat{i}_n}).\\  \nonumber
&\gamma(\hat{o}^{i_1}, \ldots, \hat{o}^{i_n})\text{: Function that swaps the encoded sequence}\\ 
&\text{sequence of observations based on $(\hat{i}_1, \ldots, \hat{i}_n)$.} \label{eq:swap}
\end{align}
Any permutation $(\hat{i}_1, \ldots, \hat{i}_n)$ of the input is represented by a substitution $\sigma$ that maps the index sequence $(i_1, \ldots, i_n)$ to the reordered sequence $(\hat{i}_1, \ldots, \hat{i}_n)$. The observations can be reordered using the function $\gamma$ that applies the substitution $\sigma$ to the observations.
The output of the action at the decoder is defined based on Multi-Agent Advantage Decomposition (Theorem 1) with the reordered observations as follows:
\begin{align}
A^{\hat{i}_{1:n}}_\pi(\hat{o}_{\text{swap}}, a^{\hat{i}_{1:n}}) = \sum_{m=1}^{n} A^{\hat{i}_{1:m}}_\pi(\hat{o}_{\text{swap}}, a^{\hat{i}_{1:m-1}}, a^{\hat{i}_m}) \label{eq:decomposition}
\end{align}
By setting an arbitrary permutation $(\hat{i}_1, \ldots, \hat{i}_n)$ as the action decision order $ao_t$, the order of the output actions follows the order of the action decisions. When updating the overall value function from the value functions of individual agents, it is possible to update the value functions serially in an arbitrary order of action decisions using Equations (\ref{eq:swap}) and (\ref{eq:decomposition}).

We propose a sequential action decision order prediction system that predicts the order in which agents take actions. The first agent is predetermined, and subsequent agents are predicted by a learner integrated into the decoder. This learner shares parameters with the action prediction component and operates as a branching subtask. Figure \ref{sequential-action-decision-order} shows this process, illustrating how the system predicts the next agent to act and determines the subsequent action decision order.

\subsection{Architecture}
Figure \ref{aoad-mat-architecture} shows the architecture of the proposed AOAD-MAT. The encoder architecture and loss function are unchanged from those of the prior MAT \cite{MAT}. For the decoder, we introduce a sequential action decision-order prediction system and design a loss function that includes subtasks.
Since the decoder (actor network) performs the order-aware policy learning, including predicting the next agent to act, it is necessary to design a loss function that converges both action and next-action agent predictions. The proposed loss functions are shown in Equations (\ref{eq:loss_encoder})--(\ref{eq:loss_decoder}).
\begin{align}\label{eq:loss_encoder}
L_{\text{Encoder}}(\phi) &= \frac{1}{Tn} \sum_{m=1}^{n} \sum_{t=1}^{T-1} \big(R(o_t, a_t) + \gamma V_{\bar{\phi}}(\hat{o}^{i_m}_{t+1})- V_{\phi}(\hat{o}^{i_m}_t)\big)^2
\end{align}
We define the state $s_t^m = (\hat{o}^{\hat{i}_{1:n}}_t, \hat{a}^{\hat{i}_{1:m-1}}_t)$ to represent the observation and action history up to the $m$-th agent at time $t$. 
The actor network outputs two probability distributions to act in the following roles:
\begin{itemize}
\item $\pi^m_a(\theta)$: Action prediction distribution. It approximates the action $a^{\hat{i}_m}$ of agent $\hat{i}_m$, which is the current agent making a decision.
\item $\pi^m_i(\theta)$: Next agent prediction distribution. It approximates $\sigma(i_{m+1})$, which determines the next agent to act in the sequence.
\end{itemize}
In other words, $\pi_a(\theta)$ predicts the optimal action for the current agent, whereas $\pi_i(\theta)$ predicts which agent should act next according to the permutation $\sigma$. This dual prediction mechanism allows our model to decide on actions and dynamically determine the order of agent decision-making.
The ratios of the probability distributions between the old and new policies are defined as follows:
\begin{align}
r_a^{m}(\theta) &= \frac{\pi_a^{m}(a^{\hat{i}_m}_t|s_t^m, \theta)}{\pi_a^{m}(a^{\hat{i}_m}_t|s_t^m, \theta_{\text{old}})} &
r_i^{m}(\theta) &= \frac{\pi_i^{m}(\hat{i}_{m+1}|s_t^m, \theta)}{\pi_i^{m}(\hat{i}_{m+1}|s_t^m, \theta_{\text{old}})}
\end{align}
The decoder's loss function in AOAD-MAT is defined as:
\begin{align}\label{eq:loss_decoder}
L_{\text{Decoder}}(\theta) = &-\frac{1}{Tn} \sum_{m=1}^{n} \sum_{t=1}^{T-1}\Big(
\min\big(r^{m}(\theta) \hat{A}_t, \text{clip}(r^{m}(\theta), 1 \pm \epsilon) \hat{A}_t\big) \Big) \\ \nonumber
&- \beta_1 H[\pi_a(\theta)] - \beta_2 H[\pi_i(\theta)]
\end{align}
where $r^{m}(\theta) = r_a^{m}(\theta) \cdot r_i^{m}(\theta)$. $H[\cdot]$ denotes the entropy, and $\beta_1, \beta_2$ are the Hyper-parameters. $\epsilon$ denotes the PPO clip parameter, which limits the extent of policy updates to ensure stable learning.

We calculate the product of $r_a^{m}(\theta)$ and $r_i^{m}(\theta)$ to obtain the overall ratio $r^{m}(\theta)$, which is then used in the advantage function. 
Taking the product of ratios has the effect of strongly promoting policy updates when they are in the same direction and strongly suppressing them when they are in different directions.
This approach differs from traditional multitask learning, where a weighted sum of loss functions is typically used. 
For comparison, we also implemented and tested the traditional multitask learning approach using the following loss function:
\begin{align}\label{eq:loss_decoder_sum}
L_{\text{Decoder}}^{\text{sum}}(\theta) &= -\frac{1}{Tn} \sum_{m=1}^{n} \sum_{t=1}^{T-1} \Big(\alpha_1 \min\big(r_a^{\hat{i}_m}(\theta) \hat{A}_t, \text{clip}(r_a^{\hat{i}_m}(\theta), 1 \pm \epsilon) \hat{A}_t\big)\\ \nonumber
&+\alpha_2 \min\big(r_i^{\hat{i}_m}(\theta) \hat{A}_t, \text{clip}(r_i^{\hat{i}_m}(\theta), 1 \pm \epsilon) \hat{A}_t\big) \Big) \\ \nonumber
&- \beta_1 H[\pi_a(\theta)] - \beta_2 H[\pi_i(\theta)]
\end{align}
where $\alpha_1$ and $\alpha_2$ denote the weighting factors of the action and next-agent prediction tasks, respectively.

Our preliminary evaluations demonstrate that using the product of these ratios leads to more balanced and stable policy updates for both tasks than the weighted sum approach\footnote{The details of the preliminary evaluations are shown in the Supplemental Material.}.
We selected this loss function because both ratios aim to optimize the same advantage function. By using their product, we ensure that policy updates consider both tasks simultaneously, thereby avoiding potential conflicts that could arise from separate gradient updates. The entropy terms $H[\pi_a(\theta)]$ and $H[\pi_i(\theta)]$ are included in both approaches to promote exploration and prevent premature convergence in the action and next-agent prediction tasks.

\begin{algorithm}[tb]
\caption{AOAD-MAT: Inference phase}\label{alg:Execution_Phase}
\begin{algorithmic}[1]
    \Require Step size $\alpha$, batch size $B$, number of agents $n$, episodes $K$, steps per episode $T$
    \State \textit{Initialize} Encoder $\phi_0$, Decoder $\theta_0$, Replay buffer $\mathcal{B}$
    \For{$k=0,1,...,K-1$}
        \For{$t=0,1,...,T-1$}
            \State \textit{Collect} observation sequence $o^{i_1}_t, ..., o^{i_n}_t$
            \State \textit{Generate} representation sequence $\hat{o}^{i_1}_t, ..., \hat{o}^{i_n}_t$
            \Comment{feeding observations to the encoder.}
            \State \textit{Input} $\hat{o}^{i_1}_t, ..., \hat{o}^{i_n}_t$ to decoder
            \For{$m=1,...,n-1$}
                \State \textit{Input} $a^{\hat{i}_0}_t, ..., a^{\hat{i}_{m-1}}_t$
                \State \textit{Infer} $\hat{i}^{m+1}$ with auto-regressive decoder
                \State \textit{Infer} $a^{\hat{i}_{m}}_t$ with the auto-regressive decoder
                \State \textit{Swap} representation sequence $\hat{o}^{i_1}_t, ..., \hat{o}^{i_n}_t$ by action decision order $ao_t$ 
            \EndFor
            \State \textit{Restore} original action sequence $a^{i_0}_t, ..., a^{i_n}_t$ by  by action decision order $ao_t$
            \State \textit{Execute} joint actions $a^{i_0}_t, ..., a^{i_n}_t$
            \State \textit{Collect} rewards $R(o_t, a_t)$
            \State \textit{Insert} $(o_t, a_t, R(o_t, a_t))$ into $\mathcal{B}$
        \EndFor
    \EndFor
\end{algorithmic}
\end{algorithm}

\subsection{Learning Algorithm}
The inference phase of the learning algorithm for AOAD-MAT is shown in Algorithm \ref{alg:Execution_Phase}. The general learning flow is categorized into inference, training and evaluation phases. The inference phase collects data from the simulation environment, and the training phase updates the policies of the model. Finally, the model performance is evaluated in the evaluation phase\footnote{The algorithm of the training phase is shown in the Supplemental Material}.

\section{Experiment}
\subsection{Experimental Setting}

\paragraph{Baselines. }
We compare the proposed AOAD-MAT model with the traditional MAT \cite{MAT} and a modified version, MAT-adjust with PPO clip that decreases from 0.05 to 0.01 and PPO-epoch tuned as the PPO clip changes from the original MAT.

\paragraph{SMAC Environment. }

\begin{figure}
\centering
\includegraphics[width=0.6\textwidth]{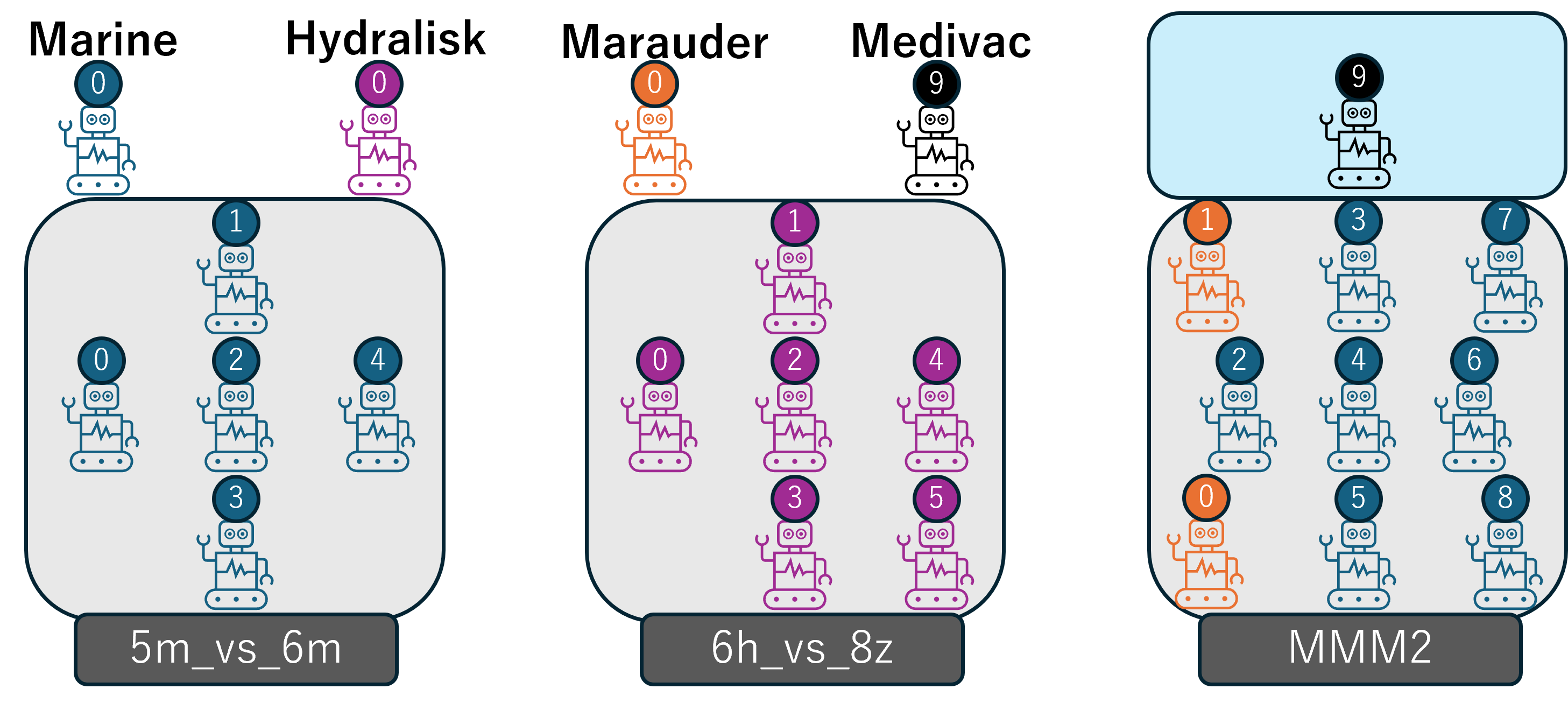}
\caption{Agent numbering in SMAC: Units are color-coded and numbered to indicate their fixed order. In MMM2, the Medivac unit is represented separately because of its aerial nature. In 5m\_vs\_6m and MMM2, the enemies are located on the right side, and they are located on the top side in 6h\_vs\_8z.}
\label{fig:agent_numbering_SMAC}
\end{figure}

SMAC is a benchmark environment for MARL based on the complex real-time strategy game StarCraft II \cite{SMAC}. We selected four challenging tasks in SMAC: 5m\_vs\_6m, 6h\_vs\_8z, MMM2, and 3s5z\_vs\_3s6z. These tasks were selected due to their high difficulty level because the original MAT methods have already achieved 100\% win rates in most other SMAC tasks.
The selected tasks include the homogeneous (5m\_vs\_6m, 6h\_vs\_8z) and heterogeneous (MMM2, 3s5z\_vs\_3s6z) unit compositions. Homogeneous tasks challenge agents with numerical disadvantages, requiring efficient attack strategies, focus fire, and kiting techniques. In contrast, heterogeneous tasks add complexity by introducing units with varying capabilities, necessitating role understanding and strategic action selection.
To investigate the influence of the lead agent, we analyzed the initial positions of allied units in 5m\_vs\_6m, 6h\_vs\_8z, and MMM2 tasks. As shown in Figure \ref{fig:agent_numbering_SMAC}, each number depicts the initial position of applied units.
The performances in SMAC tasks are presented as median win rate and 25\%-75\% percentiles over 32 test episodes across five independent runs.

\begin{figure}[tb]
\centering
\includegraphics[width=0.55\textwidth]{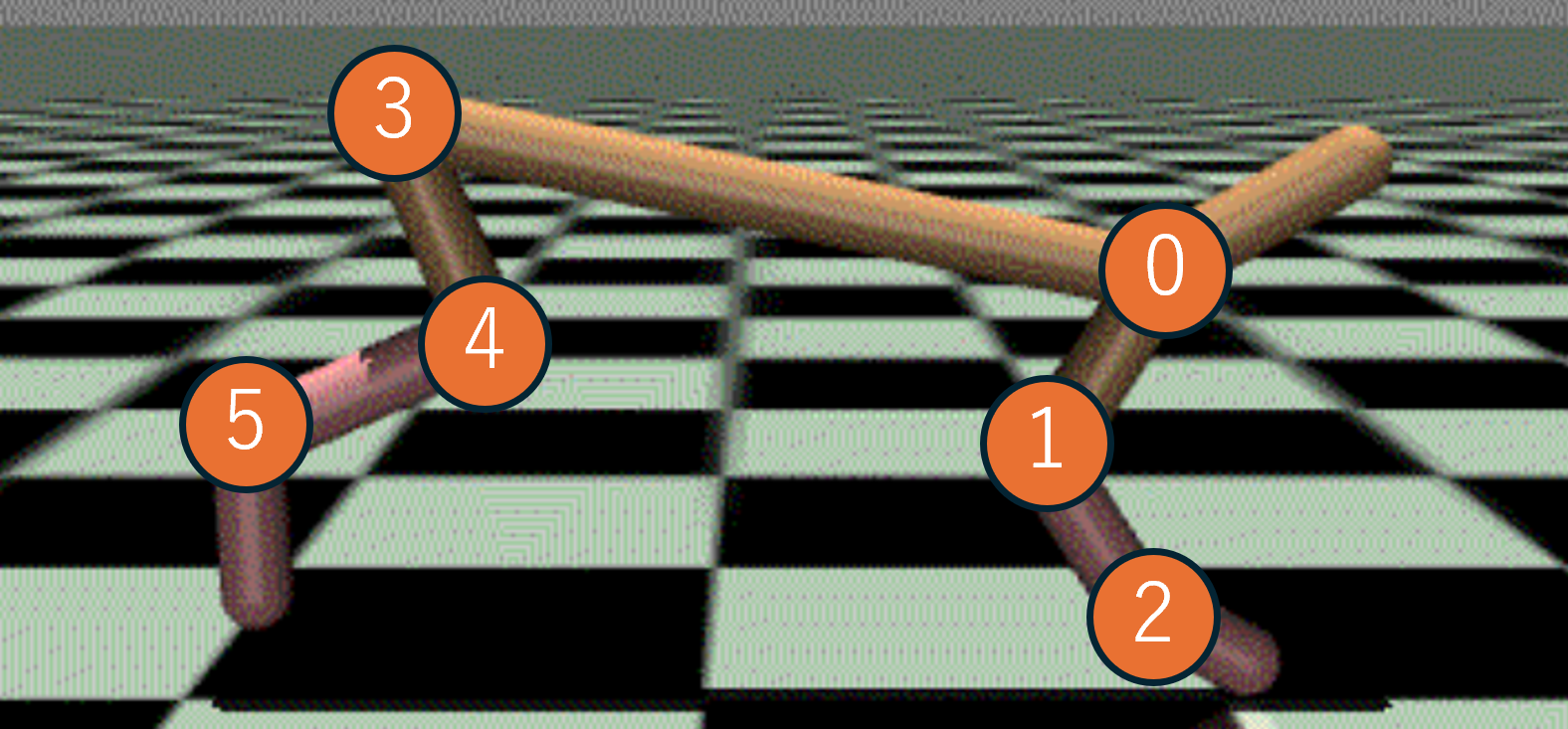}
\caption{Agent numbering in HalfCheetah (6×1): Each number of the HalfCheetah's joints shows the fixed order.}
\label{fig:agent_numbering_MuJoCo}
\end{figure}

\paragraph{MA-MuJoCo Environment.}
MA-MuJoCo extends the MuJoCo physics simulation engine, which is widely used for single-agent reinforcement learning, to MARL scenarios \cite{MA-MuJoCo}. We focus on the HalfCheetah (6×1) task, which requires controlling six joints of the HalfCheetah robot to maximize its forward velocity. This task presents a continuous action space in contrast to SMAC's discrete actions, and demands complex cooperative behavior among the joints numbered (Figure \ref{fig:agent_numbering_MuJoCo}).
We evaluated the mean of the average reward and 95\% confidence intervals over five test episodes for five independent runs.

\begin{table*}
\centering
\caption{Median and standard deviation of win rates in SMAC and mean and standard deviation of average episode rewards in MuJoCo and the number of execution steps in evaluation phase}\label{Tab:Median}
\begin{tabular}{@{}lll|ccc|c@{}}
Benchmark & Task & Difficulty & AOAD-MAT & MAT-adjust & MAT & Steps\\ \midrule
SMAC & 5m\_vs\_6m & Hard & \textbf{100.0}(1.4) & \textbf{100.0}(1.4) & 96.9(1.4) & $1 \times 10^8$ \\
 & MMM2 & Hard+ & \textbf{100.0}(1.4) & \textbf{100.0}(1.4) & \textbf{100.0}(1.4) & $5 \times 10^7$  \\
 & 6h\_vs\_8z & Hard+ & \textbf{100.0}(0.0) & \textbf{100.0}(0.0) & \textbf{100.0}(0.0) & $1.5 \times 10^7$  \\ 
 & 3s5z vs 3a6z & Hard+ & \textbf{100.0}(1.7) & 96.9(3.4) & \textbf{100.0}(0.0) & $5\times 10^7$  \\ 
MA-MuJoCo & HalfCheetah (6×1) & Expert & \textbf{13686}(1286) & 12778(1090) & 11970(256) &  $1 \times 10^8$ \\ \bottomrule
\end{tabular}
\end{table*}

\subsection{Experimental Result}
\subsubsection{Performance on SMAC Tasks}
\begin{figure}
    \centering
    \begin{tabular}{cc}
      \begin{minipage}[t]{0.45\textwidth} 
        \centering
        \includegraphics[width=\linewidth]{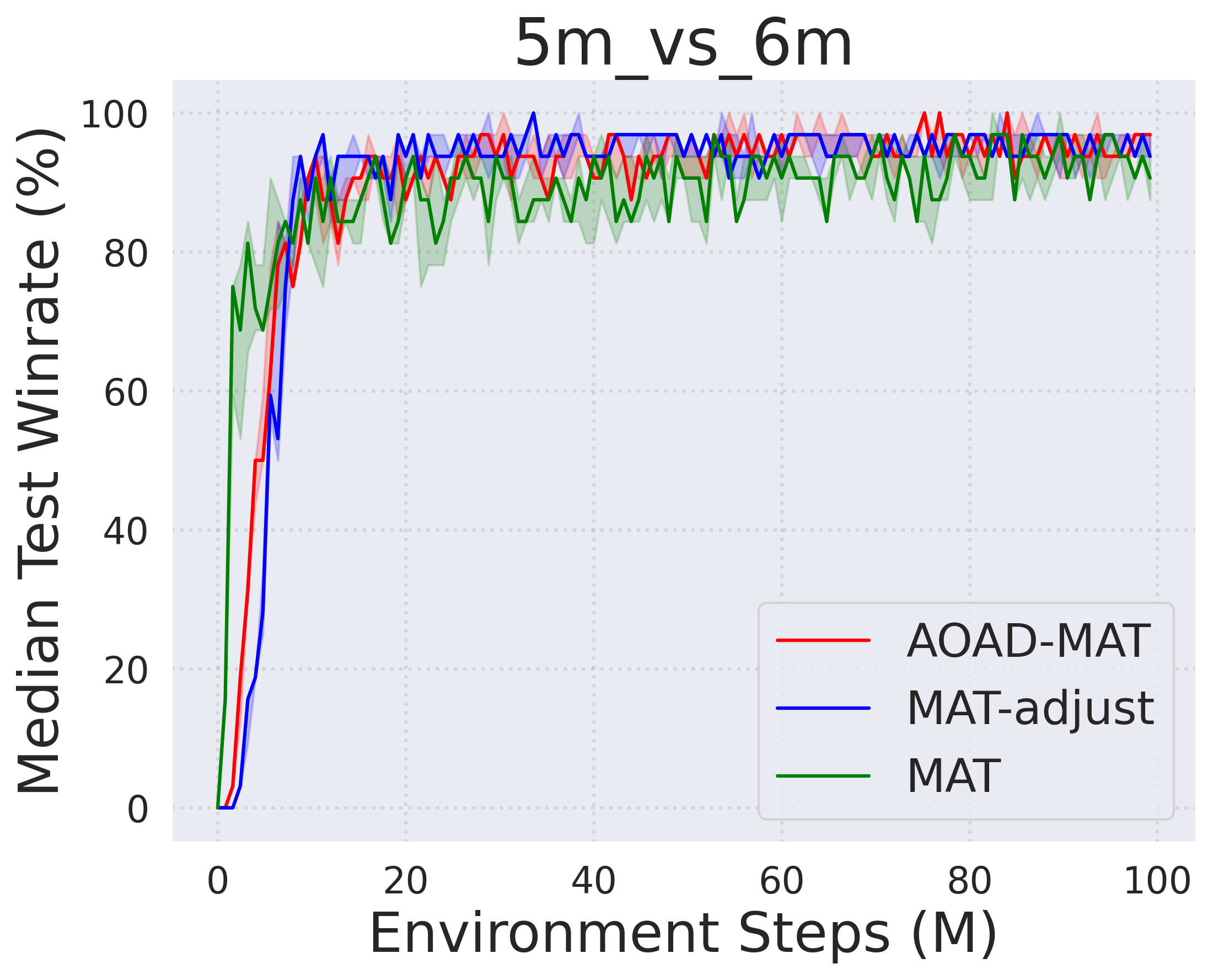} 
      \end{minipage} &
      \begin{minipage}[t]{0.45\textwidth} 
        \centering
        \includegraphics[width=\linewidth]{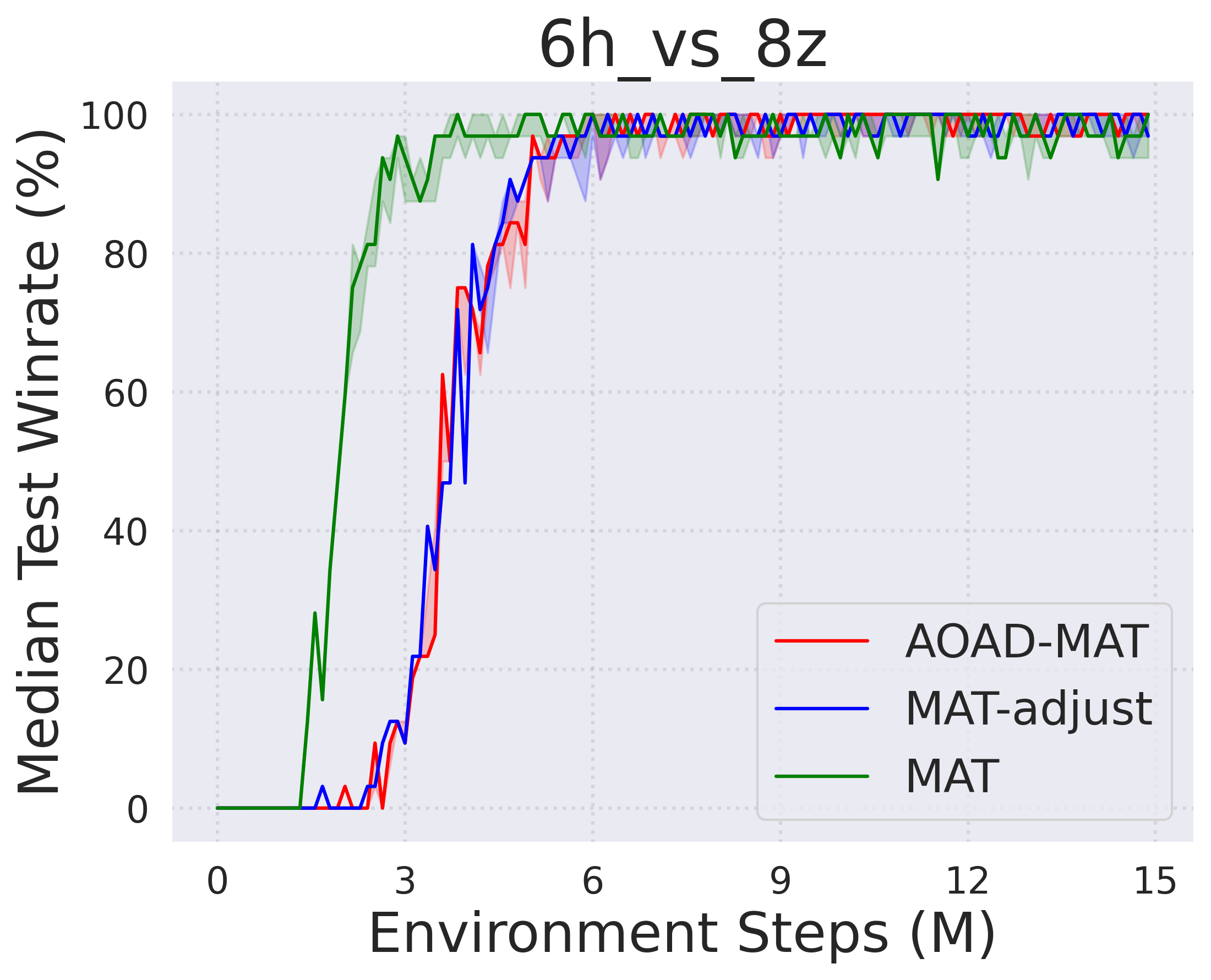}
      \end{minipage} \\ \\ 

      \begin{minipage}[t]{0.45\textwidth} 
        \centering
        \includegraphics[width=\linewidth]{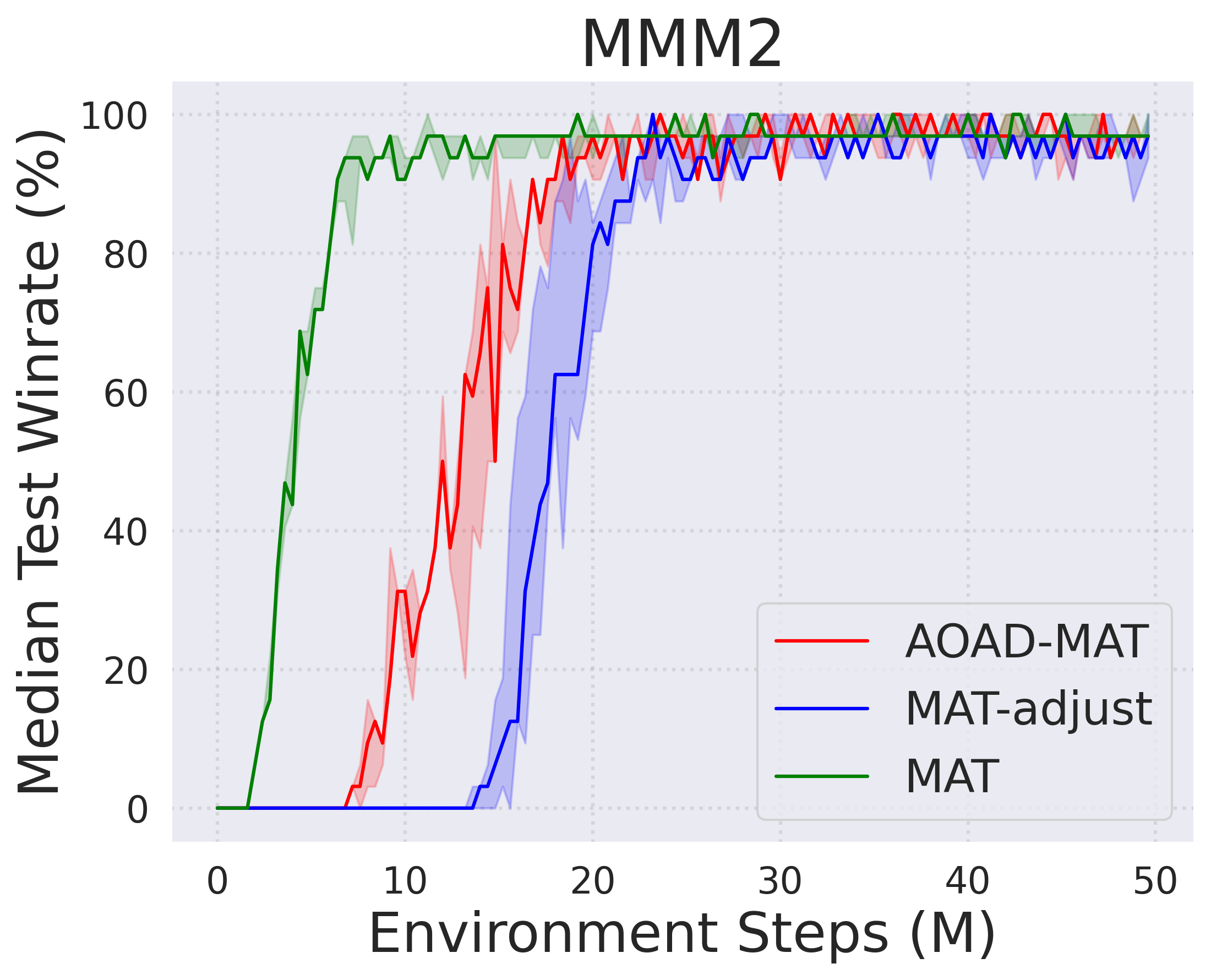}
      \end{minipage} &
      \begin{minipage}[t]{0.45\textwidth} 
        \centering
        \includegraphics[width=\linewidth]{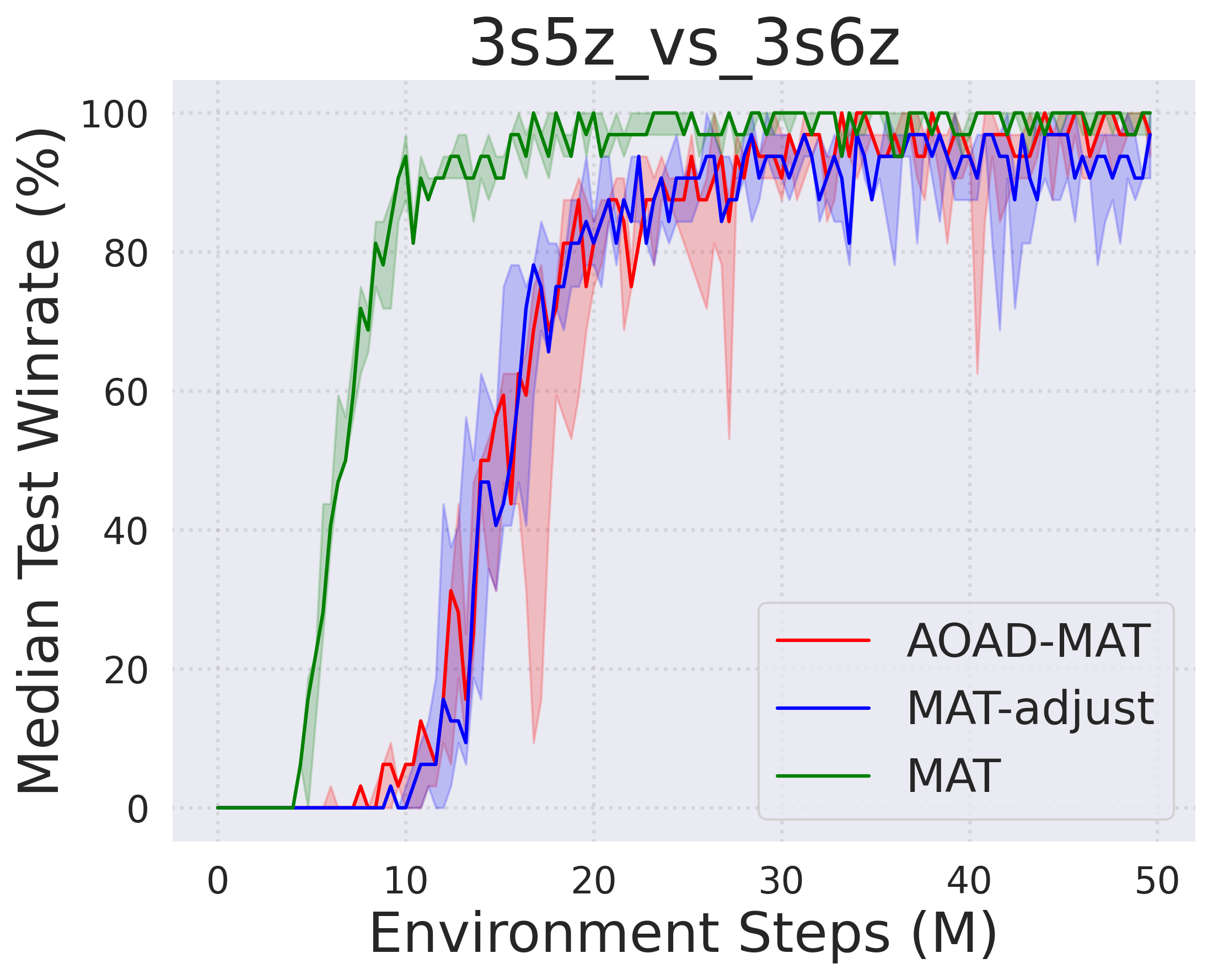}
      \end{minipage}
    \end{tabular}
    \caption{Median win rate (\%) in the SMAC task among AOAD-MAT, MAT-adjust, and MAT in the evaluation phase.}
    \label{fig:median-win-rate}
\end{figure}

We compared  the proposed AOAD-MAT with MAT and MAT-adjust (with reduced PPO clip value) on four SMAC scenarios: 5m\_vs\_6m, 6h\_vs\_8z, MMM2, and 3s5z\_vs\_3s6z. Since all models achieved 100\% win rates (Table \ref{Tab:Median}), we analyze top $n\%$ step performance for differentiation in Table \ref{tab:top_n}.
Table \ref{tab:top_n} shows that AOAD-MAT consistently achieves the highest performance across tasks and percentiles, notably outperforming others in MMM2. Although MAT showed faster convergence (Figure \ref{fig:median-win-rate}), the proposed AOAD-MAT achieves the highest win rate after 30 M steps.
The consistently high win rates observed across many tasks suggest that the improvement is not simply due to longer training time; rather, the action-order-aware design of AOAD-MAT contributes to more stable policy updates.

Many MARL methods show high performance while getting worse because of the suboptimal PPO clip selection. MAT-adjust shows that the lower PPO clip values lead to slower convergence but more stable improvements in difficult tasks, especially in homogeneous scenarios with fewer agents (5m\_vs\_6m, 6h\_vs\_8z).
However, adjusting the PPO clip value alone in complex heterogeneous tasks (MMM2, 3s5z\_vs\_3s6z) is insufficient. The superior performance of the proposed AOAD-MAT in all scenarios means that its action ordering strategy provides benefits beyond PPO clip adjustment. 

\subsubsection{Performance on MA-MuJoCo Tasks} 
Table \ref{Tab:Median} demonstrates that the proposed AOAD-MAT achieves approximately 10\% improvement in median reward compared to baselines with a substantial increase in the 95\% confidence interval's upper bound. In other words, the proposed AOAD-MAT achieved higher peak performance than the baseline methods.
Additionally, the proposed AOAD-MAT's adaptive ordering strategy achieves more stable learning and higher peak performance than fixed or random orderings in Figure \ref{fig:sort-result}(d). The monotonic increase in performance over training steps demonstrates AOAD-MAT's learning stability, which is crucial in continuous control tasks. These results demonstrate the effectiveness of the proposed AOAD-MAT in complex multi-agent environments.

\begin{figure}
  \centering
  \subfigure[Average rewards]{\label{reward}\includegraphics[width=0.4\linewidth]{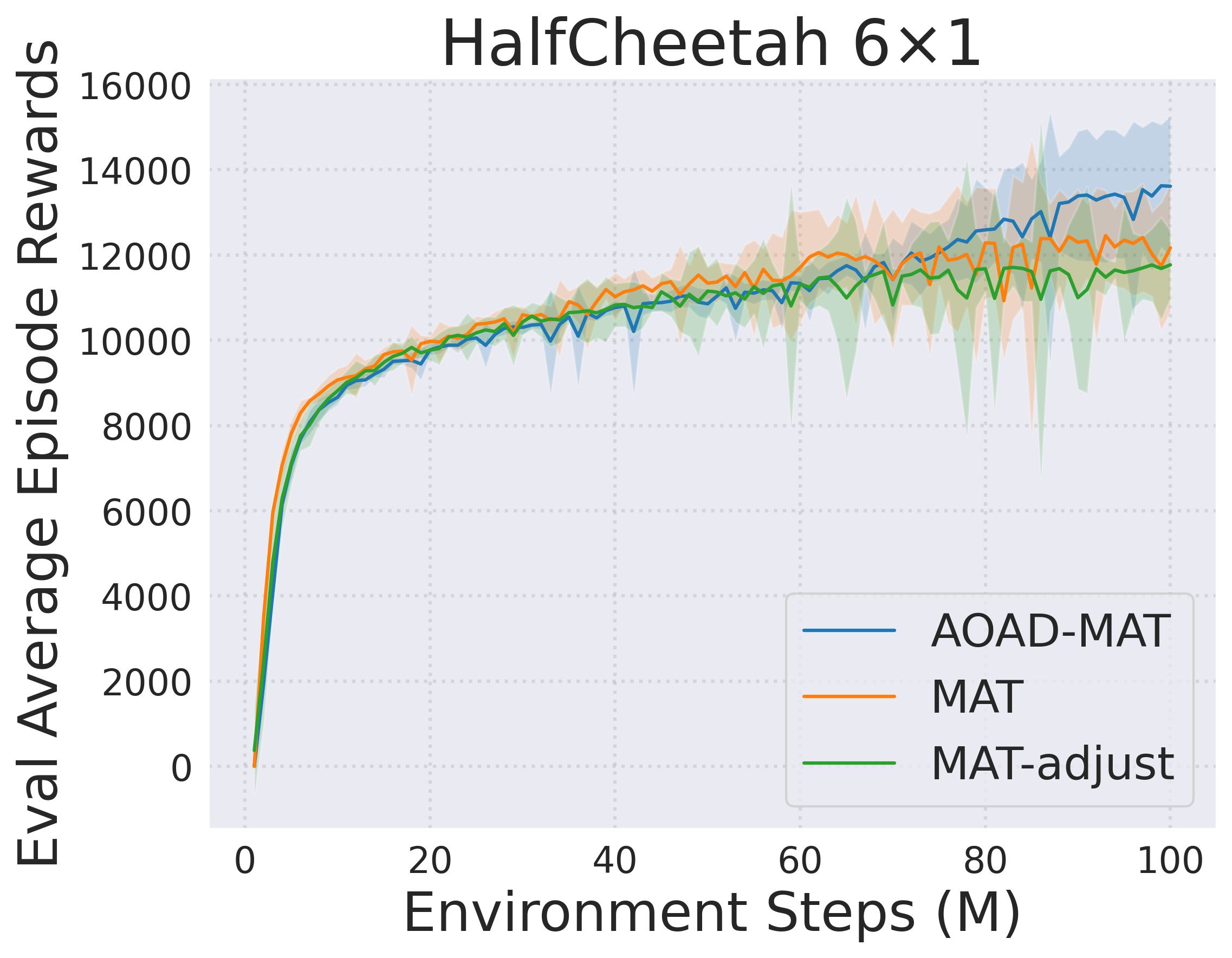}}
  \subfigure[Entropy progression]{\label{entropy}\includegraphics[width=0.4\linewidth]{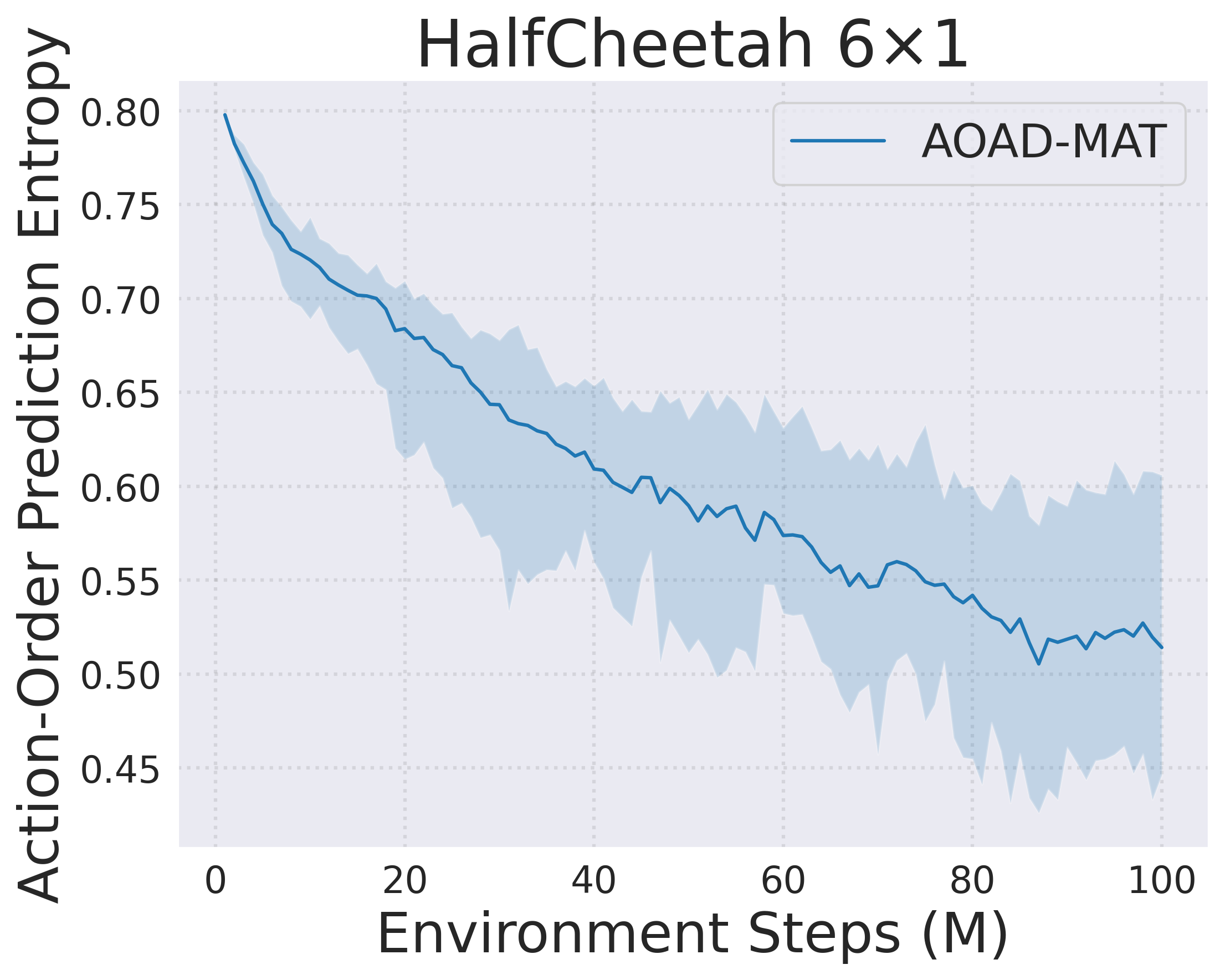}}
    \caption{
    (a): Average rewards in MuJoCo task among AOAD-MAT, MAT-adjust, and MAT in the evaluation phase. (b): Entropy progression for predicting action decisions order of AOAD-MAT}
  \label{fig:entropy}
\end{figure}

\begin{figure}
    \centering
    \begin{tabular}{cc}
      \begin{minipage}[t]{0.4\textwidth} 
        \centering
        \includegraphics[width=\linewidth]{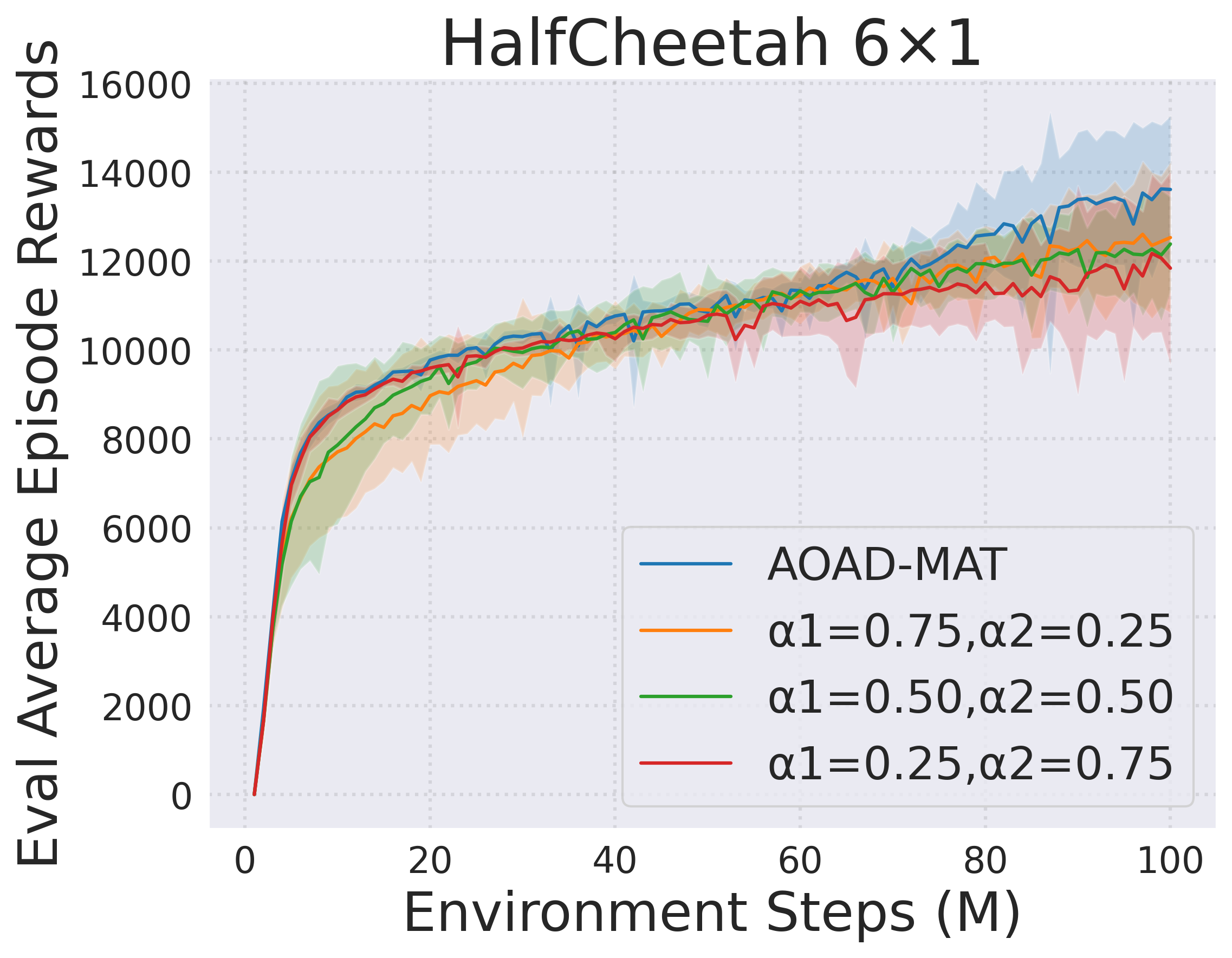} 
        (a) Average reward (MA-MuJoCo)
      \end{minipage} &
      \begin{minipage}[t]{0.4\textwidth} 
        \centering
        \includegraphics[width=\linewidth]{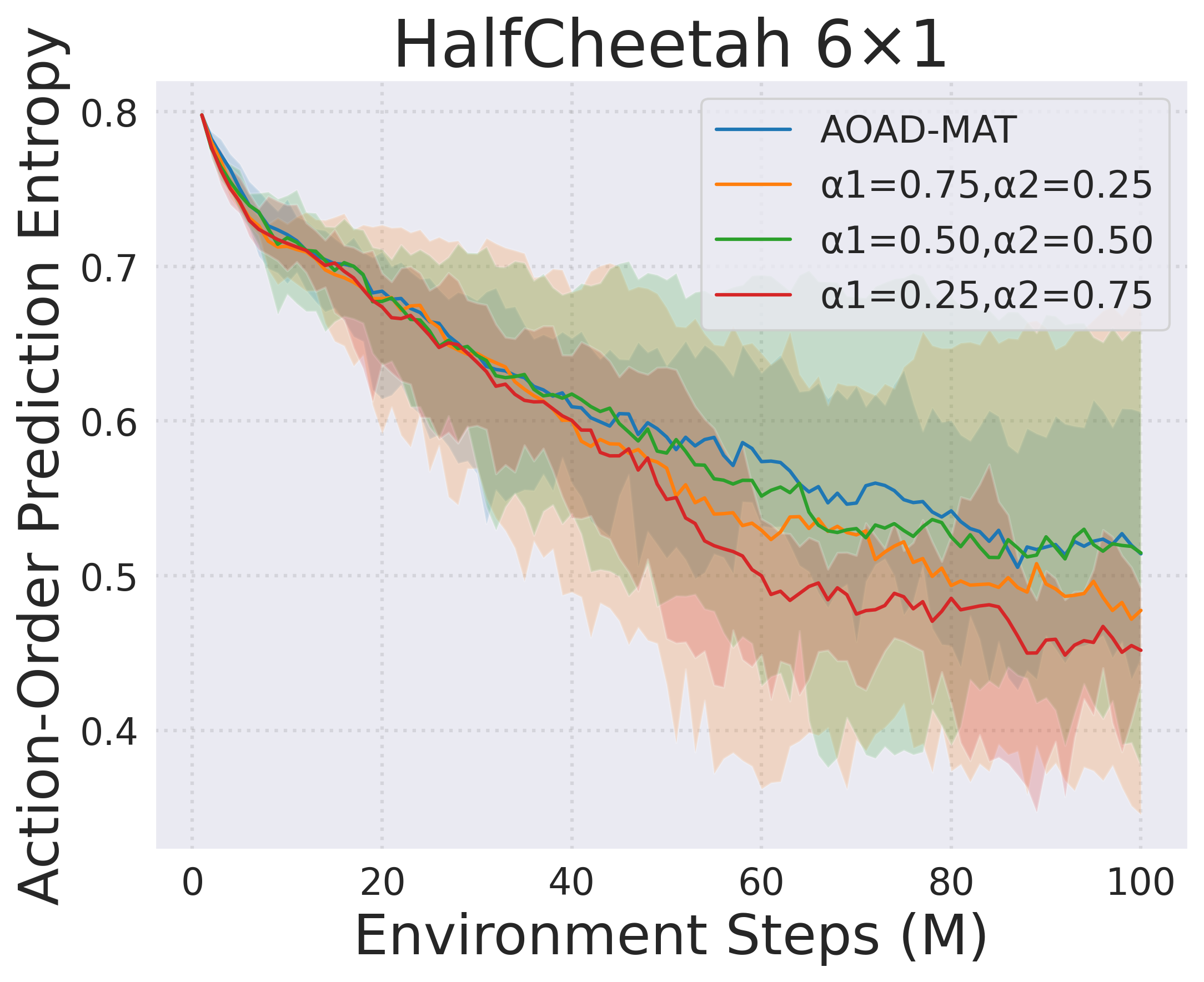}
        (b) Entropy progression (MA-MuJoCo)
      \end{minipage} \\ \\ 

      \begin{minipage}[t]{0.4\textwidth} 
        \centering
        \includegraphics[width=\linewidth]{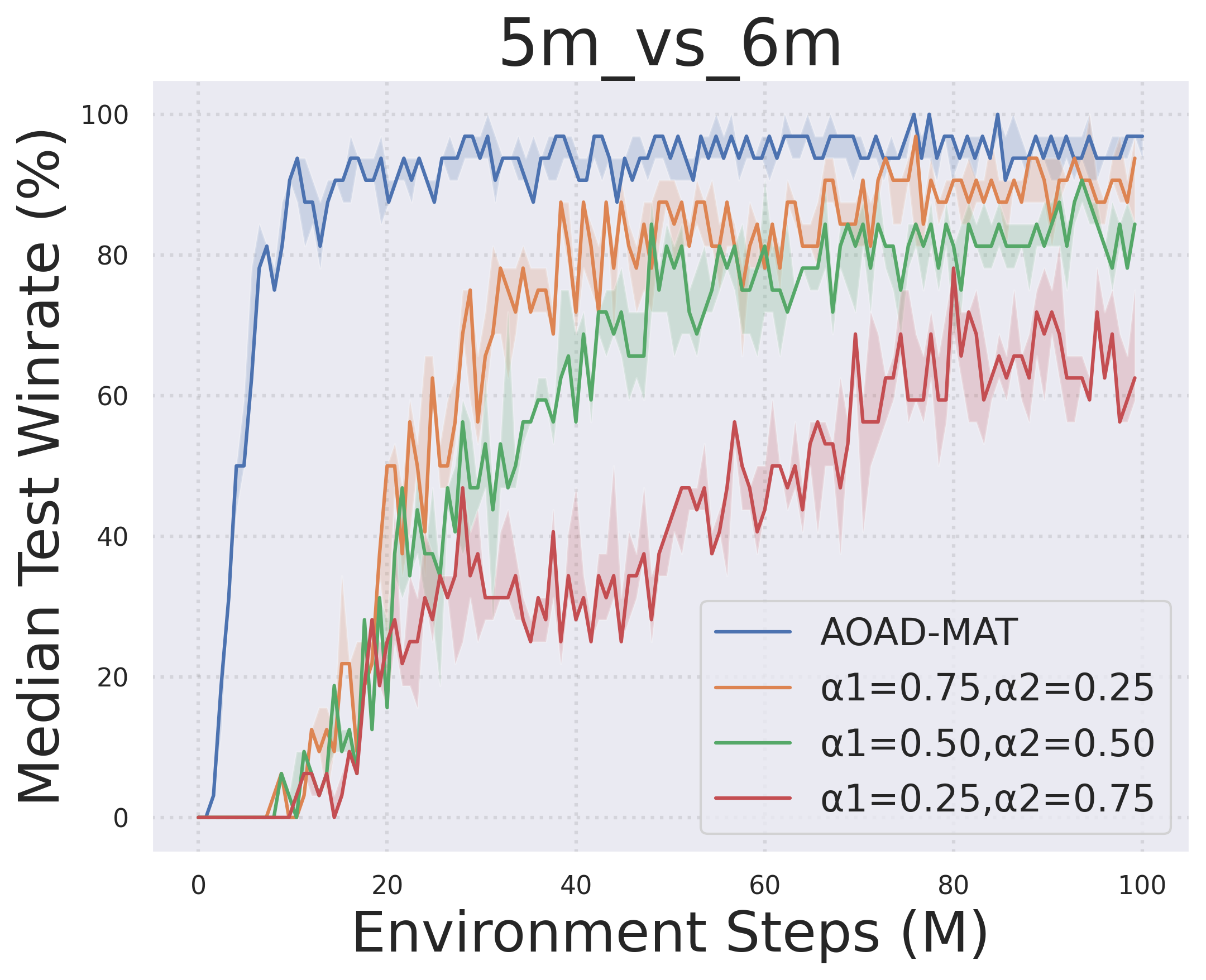}
        (c) Average reward (SMAC)
      \end{minipage} &
      \begin{minipage}[t]{0.4\textwidth} 
        \centering
        \includegraphics[width=\linewidth]{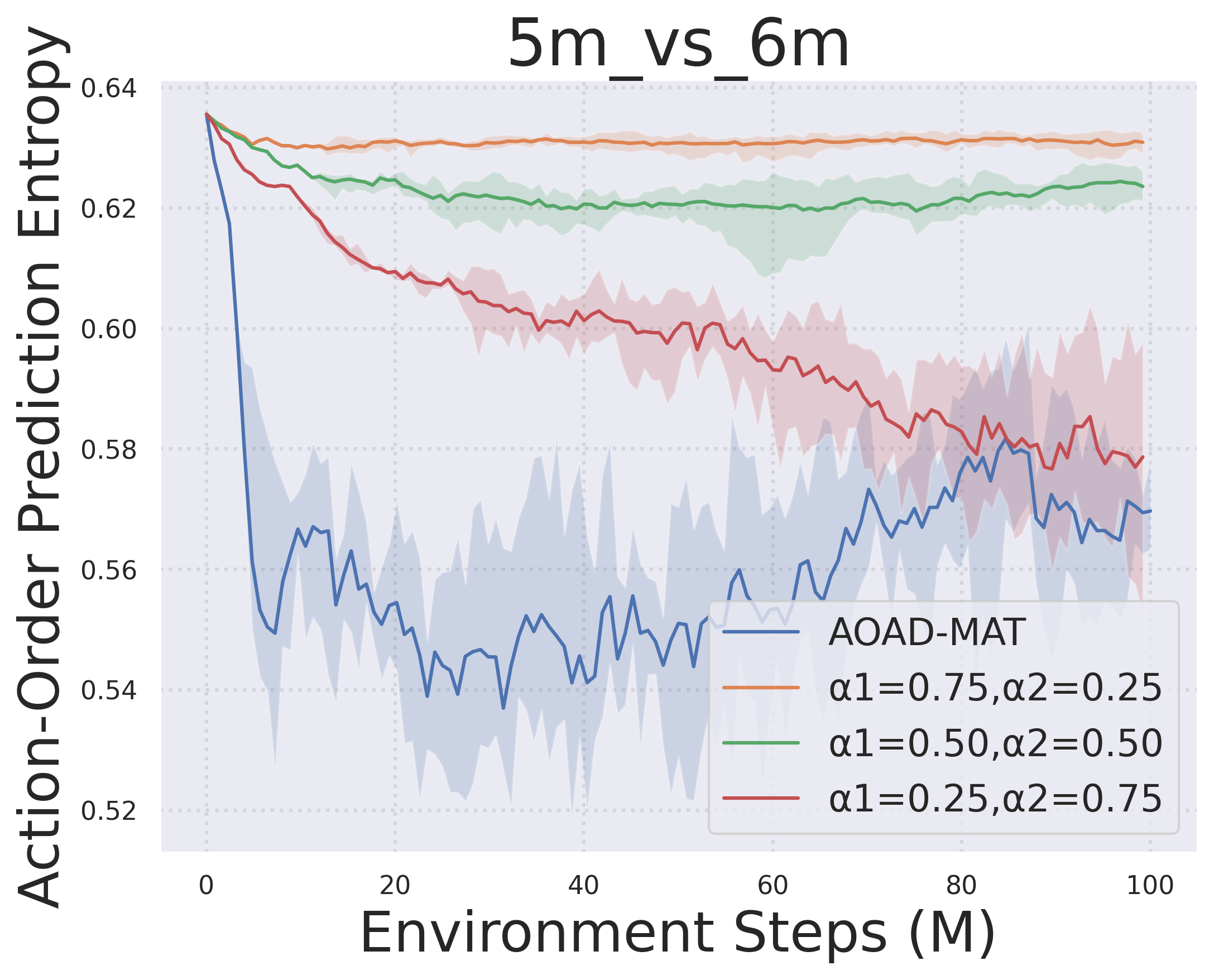}
        (d) Entropy progression (SMAC)
      \end{minipage}
    \end{tabular}
    \caption{Average reward and entropy progression for predicting action decisions order when changing the Actor's loss function in SMAC's 5m\_vs\_6m scenario and MA-MuJoCo's HalfCheetah scenario. $\alpha_1$ and $\alpha_2$ are weights in the loss function (\ref{eq:loss_decoder_sum}).}
    \label{fig:multitask}
\end{figure}

\begin{table}
\centering
\caption{Average of top $n$ steps in median win rate}\label{tab:top_n}
\begin{tabular}{@{}ll|cccc@{}}
Task & Top & AOAD-MAT & MAT-adujust & MAT\\ \midrule
5m\_vs\_6m & 5\% & \textbf{98.2}& 97.3 & 96.8\\
& 25\% & \textbf{97.2} & 97.0 & 94.7\\
& 50\% & 95.8 & \textbf{96.5} & 93.2\\ \midrule
MMM2 & 5\% & \textbf{100.0}& 98.7 & \textbf{100.0}\\
& 25\% & \textbf{99.1} & 97.3 & 97.9\\
& 50\% & \textbf{98.0} & 95.8 & 97.4\\ \midrule
6h\_vs\_8z & 5\% & \textbf{100.0}& \textbf{100.0} & \textbf{100.0}\\
& 25\% & \textbf{100.0} & \textbf{100.0} &\textbf{100.0}\\
& 50\% & \textbf{99.6} & 98.8 & 98.7\\ \midrule
3s5z\_vs\_3s6z& 5\% & \textbf{100.0} & 96.9 & \textbf{100.0}\\
& 25\% & 97.9 & 95.2 & \textbf{100.0}\\
& 50\% & 95.3 & 93.1 & \textbf{99.2}\\ \bottomrule
\end{tabular}
\end{table}

\subsubsection{Effectiveness of Agent Action Order}
Figure \ref{fig:sort-result}(d) shows that the performance of the proposed AOAD-MAT improved after 80 M steps, although the original MAT improved until 80 M steps. 
Furthermore, as shown in Figure \ref{fig:entropy}, one possible explanation for the performance gain around 80 M steps is the convergence of the entropy progression for predicting the action decisions order of AOAD-MAT. The entropy gradually decreases up to around 80 M steps, and the performance gap with the baseline methods becomes more apparent after this point. This suggests that the improvement is not simply due to extended training time, but rather the result of achieving an optimal agent action order that stabilizes policy updates.
This means that AOAD-MAT's adaptive action order becomes particularly effective in later stages of learning due to its ability to navigate the vast exploration space of MA-MuJoCo environments more efficiently.
The advantage function maximized by the actor network typically represents the value of individual agent actions. However, the Multi-Agent Advantage Decomposition used in previous approaches (like MAT) calculated inter-agent advantage differences in a fixed order. AOAD-MAT's dynamic ordering allows for a more flexible assessment of each agent's contribution, thereby enabling more adaptive and efficient exploration strategies.

Compared with the random order strategy, the fixed order slightly outperforms the random order. This means that the difference is not the diversity of the order itself, such as random order, but the quality of the predicted order. AOAD-MAT's ability to predict high-quality action orders plays a crucial role in improving exploration performance.
The reverse agent numbering consistently outperforms ascending order across all steps among fixed order strategies. As shown in Figure \ref{fig:agent_numbering_MuJoCo}, the reverse agent numbering control from the rear foot to the front foot is better than that from the front foot. However, the proposed AOAD-MAT demonstrates superior performance even though it is a good sequence heuristically compared to adaptive ordering.

\subsubsection{Influence of Lead Agent Selection}
\begin{figure}
    \centering
    \begin{tabular}{cc}
      \begin{minipage}[t]{0.4\textwidth} 
        \centering
        \includegraphics[width=\linewidth]{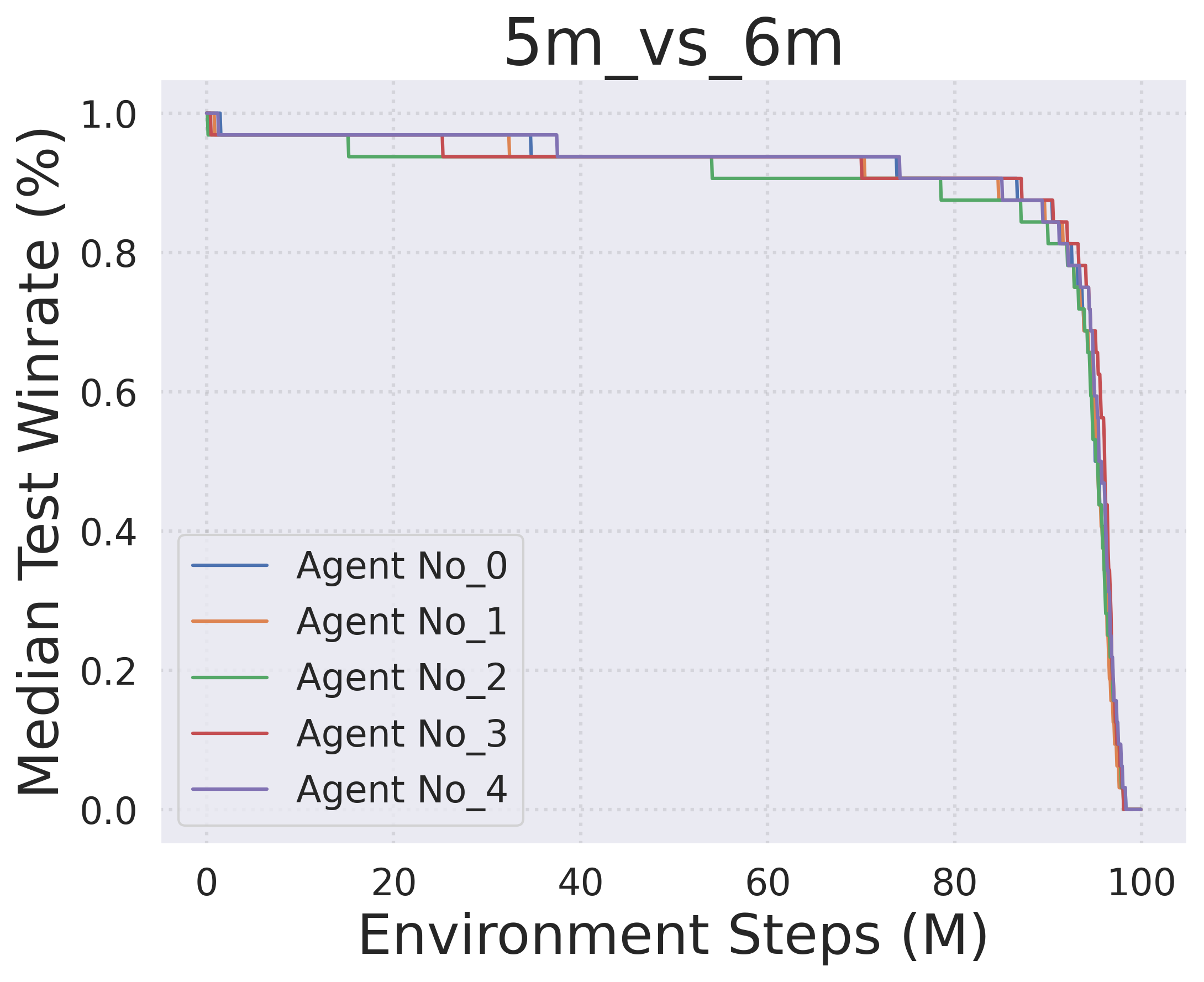} 
        (a) 5m\_vs\_6m
      \end{minipage} &
      \begin{minipage}[t]{0.4\textwidth} 
        \centering
        \includegraphics[width=\linewidth]{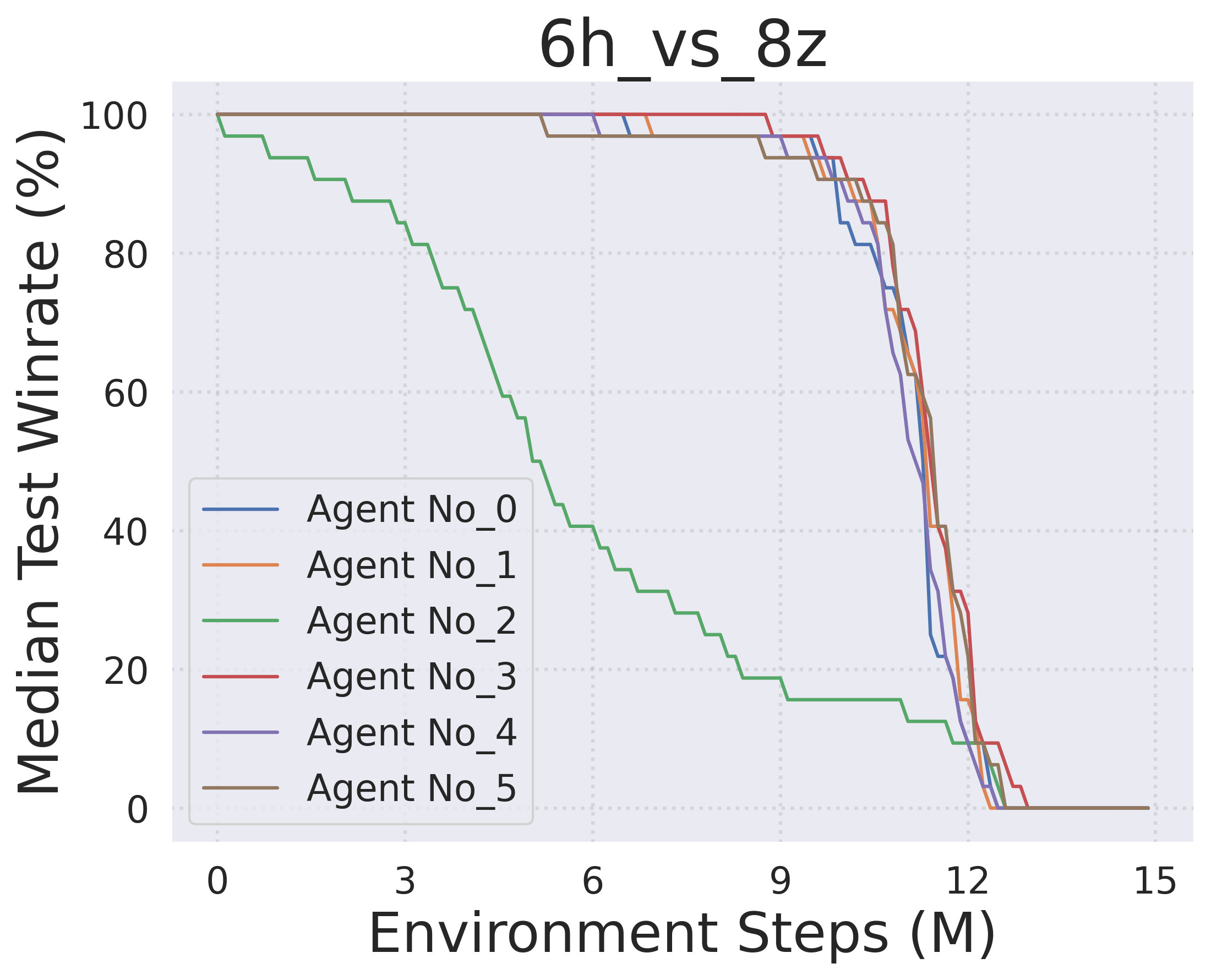}
        (b) 6h\_vs\_8z
      \end{minipage} \\ 

      \vspace{1em} \\ 

      \begin{minipage}[t]{0.4\textwidth} 
        \centering
        \includegraphics[width=\linewidth]{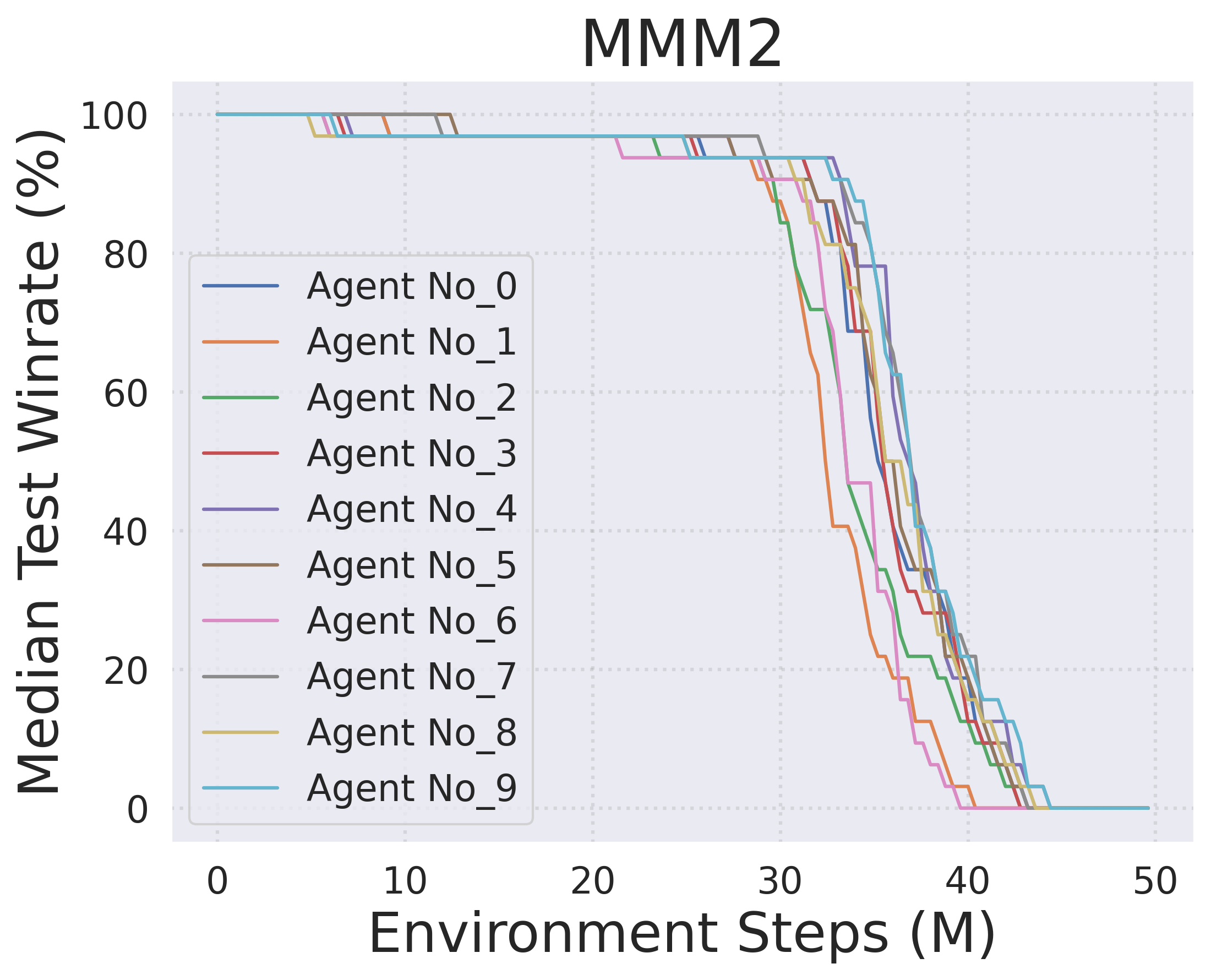}
        (c) MMM2
      \end{minipage} &
      \begin{minipage}[t]{0.4\textwidth} 
        \centering
        \includegraphics[width=\linewidth]{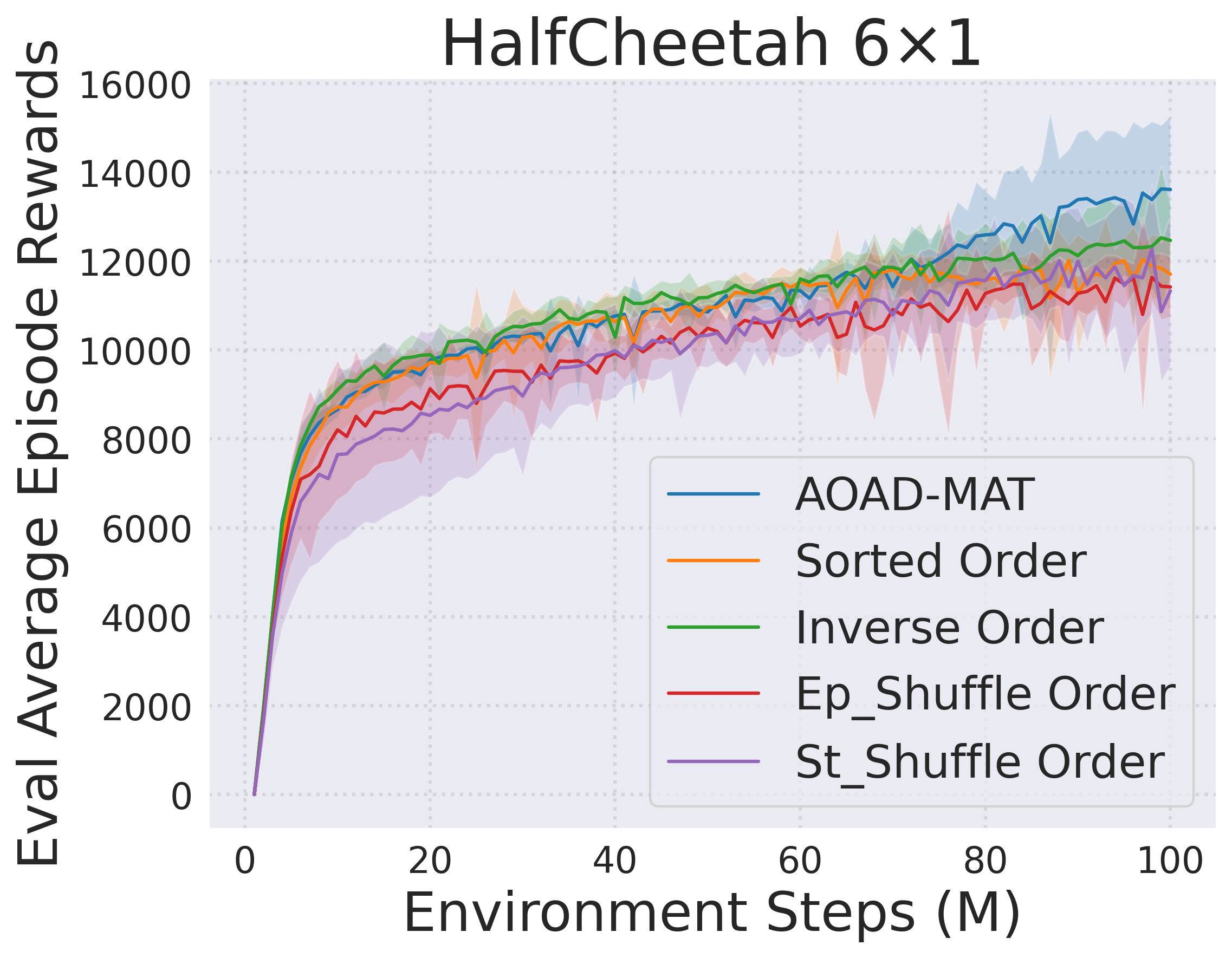}
        (d) HalfCheetah Task
      \end{minipage} \\
    \end{tabular}
    \caption{(a)-(c): Median win rate in changing the first decided agent's position in the SMAC tasks. For example, ``Agent No\_0" means the agent's position 0 is decided first. (d): Average rewards in MuJoCo under different decision orders. Agent orders are based on their agent numbering (Figure \ref{fig:agent_numbering_MuJoCo}). We evaluated four decision orders: ascending ("Sorted Order"), descending ("Inverse Order"), random per episode ("Ep Shuffle Order"), and random per step ("St Shuffle Order").}
    \label{fig:sort-result}
\end{figure}

Figures \ref{fig:sort-result}(a)-(c) show the effectiveness of lead agent selection in AOAD-MAT for the scenarios 5m\_vs\_6m, 6h\_vs\_8z, and MMM2 in the SMAC task. 
The effectiveness of lead agent selection means the role of agent positioning and task structure. In the homogeneous tasks (5m\_vs\_6m and 6h\_vs\_8z), the effective selection of the central agents or those surrounded can achieve a stable win rate. 
For heterogeneous and larger-scale tasks (MMM2), the relationship between agent position and stability is less due to the agent's various roles. However, lead agent selection that achieves the optimal leadership roles can improve performances in more complex tasks with heterogeneous teams.

\subsubsection{Loss Functions in Order-Aware Policy Learning}
Figure \ref{fig:multitask} shows the average reward and entropy progress of two loss functions as shown in Equations (\ref{eq:loss_decoder}) and (\ref{eq:loss_decoder_sum}) for the order-aware policy learning in the actor network. When setting up a loss function with a weighted sum as in conventional multitask learning, $\alpha_1$ and $\alpha_2$ are balanced by focusing either on the agent's actions or on predicting the action decision order.
Figure \ref{fig:multitask}(a) and (c) show that AOAD-MAT's synergistic loss function performs better in both MA-MuJoCo and SMAC scenarios. When  $\alpha_2$ is increased, the entropy decreases in both tasks despite the lowest performance. Therefore, the convergence of the action decision order prediction does not directly influence performance. Instead, the action predictions based on the order prediction improve the performances, synergistically.

\section{Conclusion}
In this paper, we proposed the AOAD-MAT model that predicts the action decision order of agents to facilitate sequential learning. The proposed AOAD-MAT explicitly incorporated action decision sequences into its learning process, allowing the model to learn and predict the optimal order of agent actions based on Transformer-based actor-critic architecture. It also cooperates with the subtask focused on predicting the next agent to act. We conducted several experiments on the SMAC and MA-MuJoCo benchmarks to evaluate the effectiveness of the proposed method. The experimental results showed that the proposed AOAD-MAT outperformed existing MAT and other baseline methods. We highlighted the effectiveness of adjusting the agent order of action decisions order in MARL.

One possible future work is to achieve a parallel and decentralized learning method by considering a decentralized actor, which extends the method to partial observation problems.

\newpage
\bibliographystyle{splncs04}
\bibliography{mybibfile}

\begin{thebibliography}{10}
\providecommand{\url}[1]{\texttt{#1}}
\providecommand{\urlprefix}{URL }
\providecommand{\doi}[1]{https://doi.org/#1}

\bibitem{marl-book2024}
Albrecht, S.V., Christianos, F., Sch\"afer, L.: Multi-Agent Reinforcement Learning: Foundations and Modern Approaches. MIT Press (2024), \url{https://www.marl-book.com}

\bibitem{Cre}
Chang, Y.H., Ho, T., Kaelbling, L.: All learning is local: Multi-agent learning in global reward games. In: Advances in neural information processing systems. vol.~16 (2003)

\bibitem{DAG}
Chen, D., Zhang, Q.: Context-aware bayesian network actor-critic methods for cooperative multi-agent reinforcement learning. In: Proceedings of the 40th International Conference on Machine Learning. ICML'23, JMLR.org (2023)

\bibitem{H-PPO}
Fan, Z., Su, R., Zhang, W., Yu, Y.: Hybrid actor-critic reinforcement learning in parameterized action space. arXiv preprint arXiv:1903.01344  (2019)

\bibitem{pmlr-v97-jaques19a}
Jaques, N., Lazaridou, A., Hughes, E., Gulcehre, C., Ortega, P., Strouse, D., Leibo, J.Z., De~Freitas, N.: Social influence as intrinsic motivation for multi-agent deep reinforcement learning. In: Proceedings of the 36th International Conference on Machine Learning. vol.~97, pp. 3040--3049. PMLR (09--15 Jun 2019), \url{https://proceedings.mlr.press/v97/jaques19a.html}

\bibitem{FOX}
Jo, Y., Lee, S., Yeom, J., Han, S.: Fox: Formation-aware exploration in multi-agent reinforcement learning. In: Proceedings of the AAAI Conference on Artificial Intelligence. vol. 38(12), pp. 12985--12994 (2024)

\bibitem{IQL}
Kostrikov, I., Fergus, R., Tompson, J., Nachum, O.: Offline reinforcement learning with fisher divergence critic regularization. In: International Conference on Machine Learning. pp. 5774--5783. PMLR (2021)

\bibitem{HAPPO}
Kuba, J.G., Feng, X., Ding, S., Dong, H., Wang, J., Yang, Y.: Heterogeneous-agent mirror learning: A continuum of solutions to cooperative marl. arXiv preprint arXiv:2208.01682  (2022)

\bibitem{MAPG}
Kuba, J.G., Wen, M., Meng, L., Zhang, H., Mguni, D., Wang, J., et~al., Y.Y.: Settling the variance of multi-agent policy gradients. In: Advances in Neural Information Processing Systems. vol.~34, pp. 13458--13470 (2021)

\bibitem{lazaridou2017multiagent}
Lazaridou, A., Peysakhovich, A., Baroni, M.: Multi-agent cooperation and the emergence of (natural) language. In: International Conference on Learning Representations (2017), \url{https://openreview.net/forum?id=Hk8N3Sclg}

\bibitem{ACE}
Li, C., Liu, J., Zhang, Y., Wei, Y., Niu, Y., Yang, Y., Liu, Y., Ouyang, W.: Ace: Cooperative multi-agent q-learning with bidirectional action-dependency. In: Proceedings of the AAAI conference on artificial intelligence. vol. 37(7), pp. 8536--8544 (2023)

\bibitem{Lowe17}
Lowe, R., Wu, Y., Tamar, A., Harb, J., Abbeel, P., Mordatch, I.: Multi-agent actor-critic for mixed cooperative-competitive environments. In: Proceedings of the 31st International Conference on Neural Information Processing Systems. p. 6382–6393 (2017)

\bibitem{DQN}
Mnih, V., Kavukcuoglu, K., Silver, D., Graves, A., Antonoglou, I., Wierstra, D., Riedmiller, M.A.: Playing atari with deep reinforcement learning. CoRR  \textbf{abs/1312.5602} (2013), \url{http://arxiv.org/abs/1312.5602}

\bibitem{MA-MuJoCo}
Peng, B., Rashid, T., Schroeder~de Witt, C., Kamienny, P.A., Torr, P., B{\"o}hmer, W., Whiteson, S.: Facmac: Factored multi-agent centralised policy gradients. Advances in Neural Information Processing Systems  \textbf{34},  12208--12221 (2021)

\bibitem{QMIX}
Rashid, T., Samvelyan, M., De~Witt, C.S., Farquhar, G., Foerster, J., Whiteson, S.: Monotonic value function factorisation for deep multi-agent reinforcement learning. Journal of Machine Learning Research  \textbf{21}(178),  1--51 (2020)

\bibitem{GCS}
Ruan, J., Du, Y., Xiong, X., Xing, D., Li, X., Meng, L., Zhang, H., Wang, J., Xu, B.: Gcs: Graph-based coordination strategy for multi-agent reinforcement learning. In: Proceedings of the 21st International Conference on Autonomous Agents and Multiagent Systems. p. 1128–1136. AAMAS '22, International Foundation for Autonomous Agents and Multiagent Systems, Richland, SC (2022)

\bibitem{SMAC}
Samvelyan, M., Rashid, T., de~Witt, C.S., Farquhar, G., Nardelli, N., Rudner, T.G.J., Hung, C.M., Torr, P.H.S., Foerster, J., Whiteson, S.: The starcraft multi-agent challenge. arXiv preprint arXiv:1902.04043  (2019)

\bibitem{TRPO}
Schulman, J., Levine, S., Abbeel, P., Jordan, M., Moritz, P.: Trust region policy optimization. In: International conference on machine learning. pp. 1889--1897. PMLR (2015)

\bibitem{PPO}
Schulman, J., Wolski, F., Dhariwal, P., Radford, A., Klimov, O.: Proximal policy optimization algorithms. CoRR  \textbf{abs/1707.06347} (2017), \url{http://arxiv.org/abs/1707.06347}

\bibitem{QTRAN}
Son, K., Kim, D., Kang, W.J., Hostallero, D.E., Yi, Y.: Qtran: Learning to factorize with transformation for cooperative multi-agent reinforcement learning. In: International conference on machine learning. pp. 5887--5896. PMLR (2019)

\bibitem{VDN}
Sunehag, P., Lever, G., Gruslys, A., Czarnecki, W.M., Zambaldi, V., Jaderberg, M., Lanctot, M., Sonnerat, N., Leibo, J.Z., Tuyls, K., Graepel, T.: Value-decomposition networks for cooperative multi-agent learning based on team reward. In: Proceedings of the 17th International Conference on Autonomous Agents and MultiAgent Systems. p. 2085–2087 (2018)

\bibitem{Rein}
Sutton, R.S., Barto, A.G.: Reinforcement learning: An introduction. MIT press (2018)

\bibitem{PG}
Sutton, R.S., McAllester, D., Singh, S., Mansour, Y.: Policy gradient methods for reinforcement learning with function approximation. In: Solla, S., Leen, T., M\"{u}ller, K. (eds.) Advances in Neural Information Processing Systems. vol.~12, pp. 1057--1063. MIT Press (1999)

\bibitem{Transformer}
Vaswani, A., Shazeer, N., Parmar, N., Uszkoreit, J., Jones, L., Gomez, A.N., Kaiser, L., Polosukhin, I.: Attention is all you need. In: Proceedings of the 31st International Conference on Neural Information Processing Systems. pp. 6000--6010 (2017)

\bibitem{a2po}
Wang, X., Tian, Z., Wan, Z., Wen, Y., Wang, J., Zhang, W.: Order matters: Agent-by-agent policy optimization. In: The Eleventh International Conference on Learning Representations (2023), \url{https://openreview.net/forum?id=Q-neeWNVv1}

\bibitem{MAT}
Wen, M., Kuba, J., Lin, R., Zhang, W., Wen, Y., Wang, J., Yang, Y.: Multi-agent reinforcement learning is a sequence modeling problem. Advances in Neural Information Processing Systems  \textbf{35},  16509--16521 (2022)

\bibitem{IPPO}
Witt, C.S.D., Gupta, T., Makoviichuk, D., Makoviychuk, V., Torr, P.H.S., Sun, M., Whiteson, S.: Is independent learning all you need in the starcraft multi-agent challenge? ArXiv  \textbf{abs/2011.09533} (2020), \url{https://api.semanticscholar.org/CorpusID:227054146}

\bibitem{CTDE}
Yang, Y., Wen, Y., Wang, J., Chen, L., Shao, K., Mguni, D., Zhang, W.: Multi-agent determinantal q-learning. In: International Conference on Machine Learning. pp. 10757--10766. PMLR (2020)

\bibitem{MAPPO}
Yu, C., Velu, A., Vinitsky, E., Gao, J., Wang, Y., Bayen, A., Wu, Y.: The surprising effectiveness of ppo in cooperative multi-agent games. Advances in Neural Information Processing Systems  \textbf{35},  24611--24624 (2022)

\end{thebibliography}





\end{document}